\documentclass[%
superscriptaddress,
nofootinbib,
nobibnotes,
twocolumn,
amsmath,amssymb,
prb,
longbibliography,
numerical,
]{revtex4-1}

\usepackage{graphicx}
\usepackage{dcolumn}
\usepackage{bm}
\usepackage[normalem]{ulem}
\usepackage{color}
\usepackage{hyperref}
\usepackage{amsmath}
\usepackage{times}

\newcommand{\bra}[1]{\left\langle #1 \right|}
\newcommand{\ket}[1]{\left| #1 \right\rangle}
\newcommand{\avg}[1]{\left\langle #1 \right\rangle}

\newcommand{\Hamiltonian}{{\cal H}} 
\newcommand{\id}{{\normalfont\hbox{1\kern-0.15em \vrule width .8pt depth-.5pt}}}

\newcommand{\pd}[2]{\frac{\partial #1}{\partial #2}}

\newcommand{\hide}[1]{}

\newcommand{\half}{\frac{1}{2}}

\newcommand{\Kappa}{{\cal K}}

\begin{document}
\title{Photocell Optimisation Using Dark State Protection}

\author{Amir Fruchtman}
\affiliation{Department  of  Materials,  University  of  Oxford,  Oxford  OX1  3PH,  United  Kingdom}
\author{Rafael G\'omez-Bombarelli}
\affiliation{Department of Chemistry and Chemical Biology, Harvard University, Cambridge, USA}
\author{Brendon W. Lovett}
\affiliation{SUPA, School of Physics and Astronomy, University of St Andrews, KY16 9SS, United Kingdom}
\author{Erik M. Gauger}\email[Corresponding author, ]{e.gauger@hw.ac.uk}
\affiliation{SUPA, Institute of Photonics and Quantum Sciences, Heriot-Watt University, EH14 4AS, United Kingdom}

\begin{abstract}
Conventional photocells suffer a fundamental efficiency threshold imposed by the principle of detailed balance, reflecting the fact that good absorbers  must necessarily also be fast emitters. This limitation can be overcome by `parking' the energy of an absorbed photon in a dark state which neither absorbs nor emits light. Here we argue that suitable dark states occur naturally as a consequence of the dipole-dipole interaction between two proximal optical dipoles for a wide range of realistic molecular dimers. We develop an intuitive model of a photocell comprising two light-absorbing molecules coupled to an idealised reaction centre, showing asymmetric dimers are capable of providing a significant enhancement of light-to-current conversion under ambient conditions. We conclude by describing a roadmap for identifying suitable molecular dimers for demonstrating this effect by screening a very large set of possible candidate molecules. 
\end{abstract}

\date{\today}

\pacs{Valid PACS appear here}
\maketitle

\section{Introduction}

The operation of a solar energy harvesting device can be enhanced by clever design of a nanoscopic, quantum mechanical system~\cite{Dorfman2012}. Though thermodynamical considerations lead to the famous Shockley-Queisser efficiency limit for classical photocell devices~\cite{Tokihiro1993}, the `detailed balance'  underlying this limit can be broken by careful use of quantum interference. In particular, by carefully tailoring the interactions between two~\cite{Creatore2013, Killoran2015} or more~\cite{Zhang2015, Higgins2015} idealized and identical two-level energy absorbers, it is possible to prevent the re-emission of absorbed light by arranging that excitations end up in `dark' -- {\it i.e.} optically inaccessible -- states. This allows the energy to be dissipated across a target load, rather than dissipated via spontaneous emission.

It is conjectured that nature already exploits quantum-mechanical properties in order to increase the light-harvesting efficiency of photosynthesis~\cite{Mukai1999}. The most well studied system in this context is the FMO complex~\cite{Fenna1975}, which connects the antenna to the reaction centre in the light harvesting apparatus of green sulfur bacteria. It consists of seven (or eight~\cite{doi:10.1021/jz201259v}) bacteriochlorophyll (BChl a) molecules that are held in place by a messy protein scaffold, and surrounded by water at room temperature, resulting in non-identical BChl excitation energies. True quantum effects may seem unlikely in the `hot and wet' conditions of such systems. However, the observation of quantum coherent beats in experimentally measured two-dimensional electronic spectroscopy suggests otherwise~\cite{Adolphs2006,Ishizaki2009,Engel2007,Panitchayangkoon2010,Cheng2009, Kreisbeck2012,Lambert2012}.

A good definition of the term efficiency is key to quantifying a quantum advantage. One such measure is the energy transfer efficiency, {\it i.e.}~the probability of an excitation reaching the target electron acceptor after starting from a spatially localised state~\cite{Ritz2001,Olaya-Castro2008,Shabani2012}, but this does not capture all aspects of the process. An alternative approach is placing a system between two electrodes, and measuring the current through them\cite{Smirnov2009,Einax2011,Einax2013,Ajisaka2015}. Dorfman \emph{et al.} proposed a different canonical measure~\cite{Dorfman2012}: They consider the entire cycle, from absorbing a photon to extracting work, as a quantum heat engine (QHE). Procedurally, they abstract the electron acceptor to become a two-level `trap', in which transferred electrons `fill' the excited state before the action of driving a load resistor is mimicked by decay to the lower level of the trap. This gives a straightforward way of defining the power and the efficiency of the heat engine. In this picture, Fano interference may boost the photocurrent by 27\% over that of a classical cell. Subsequently, Creatore \emph{et al.}~\cite{Creatore2013} reported an efficiency gain of 35\% by introducing the different effect of {\it dark-state protection} using two identical dipole-coupled emitters. Further gains become possible for more than two chromophores~\cite{Zhang2015,Higgins2015}.

\begin{figure}
\centering
\includegraphics[width=\linewidth]{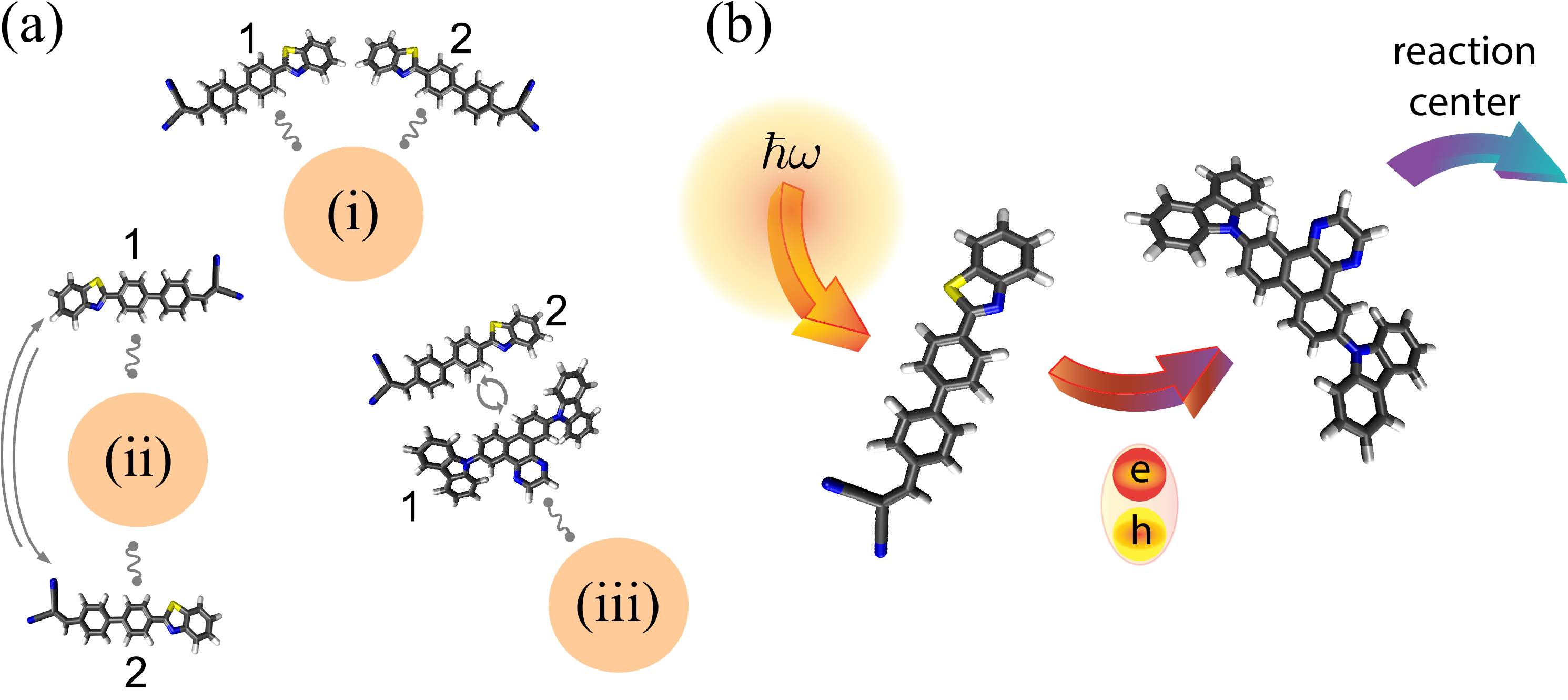}
\caption{
(a) Schematic of the different models: (i) independent, (ii) symmetric, and (iii) asymmetric, depicted with molecules from row F in Table \ref{pairs table} of Appendix \ref{Appendix A}. (In)coherent coupling is denoted by (wiggly) rounded arrows and the circle denotes the reaction center. (b) Energy flow through asymmetric model: the donor chromophore with strong optical oscillator strength absorbs a photon and transfers it to its darker partner via F\"orster transfer. Delocalisation across the dimer of the relevant quantum eigenstates is not depicted.}
\label{system glossy}
\label{models comparison}
\end{figure} 

\section{Model}

In this Letter, we use the QHE framework to determine whether a quantum advantage is achievable in non-idealized situations typical of real devices and conditions more closely resembling the photosynthetic apparatus. In particular, we consider a light harvesting device where the two constituent chromophores are not identical. Surprisingly, we will find that under realistic constraints an `asymmetric' dimer may even significantly outperform previously studied systems. 
Presenting several example molecules that would be highly efficient light harvesters according to our model, we argue that the number of conceivable molecular dimers with a quantum advantage is vast.

We now introduce a general framework that will allow us to define three specific models shortly. We consider two, generally different, light-absorbing molecules with dipolar coupling. There is a further coupling to an abstracted reaction center `trap', modelled as a two level system $\ket{\alpha}$, $\ket{\beta}$ with corresponding energies $\epsilon_\alpha, \epsilon_\beta$, following Ref.~\onlinecite{Dorfman2012}. An excitation absorbed by the molecules can be incoherently transferred into the reaction centre via a phonon-assisted process. In what follows we restrict the dynamics to the subspace with one or zero excitations across the entire system: this approximation is valid since in realistic configurations inspired by natural photosynthetic systems, the average excitation number is very small ($N \approx 0.02$ for an energy gap of $2$ eV and sunlight temperature of 6000~K). Since photoexcitation is very rare, once captured it is paramount to prevent spontaneous emission back into the environment. Hence, access to a dark state, {\it i.e.}~a state that is decoupled from the photon field and thus not susceptible to spontaneous emission decay, can enhance the engine's efficiency. By Kasha's rule re-emission of absorbed photons will be dominated by the lowest excited state, motivating our approach of only considering a single excited level per chromophore.

\begin{figure}
\centering
\includegraphics[width=\linewidth]{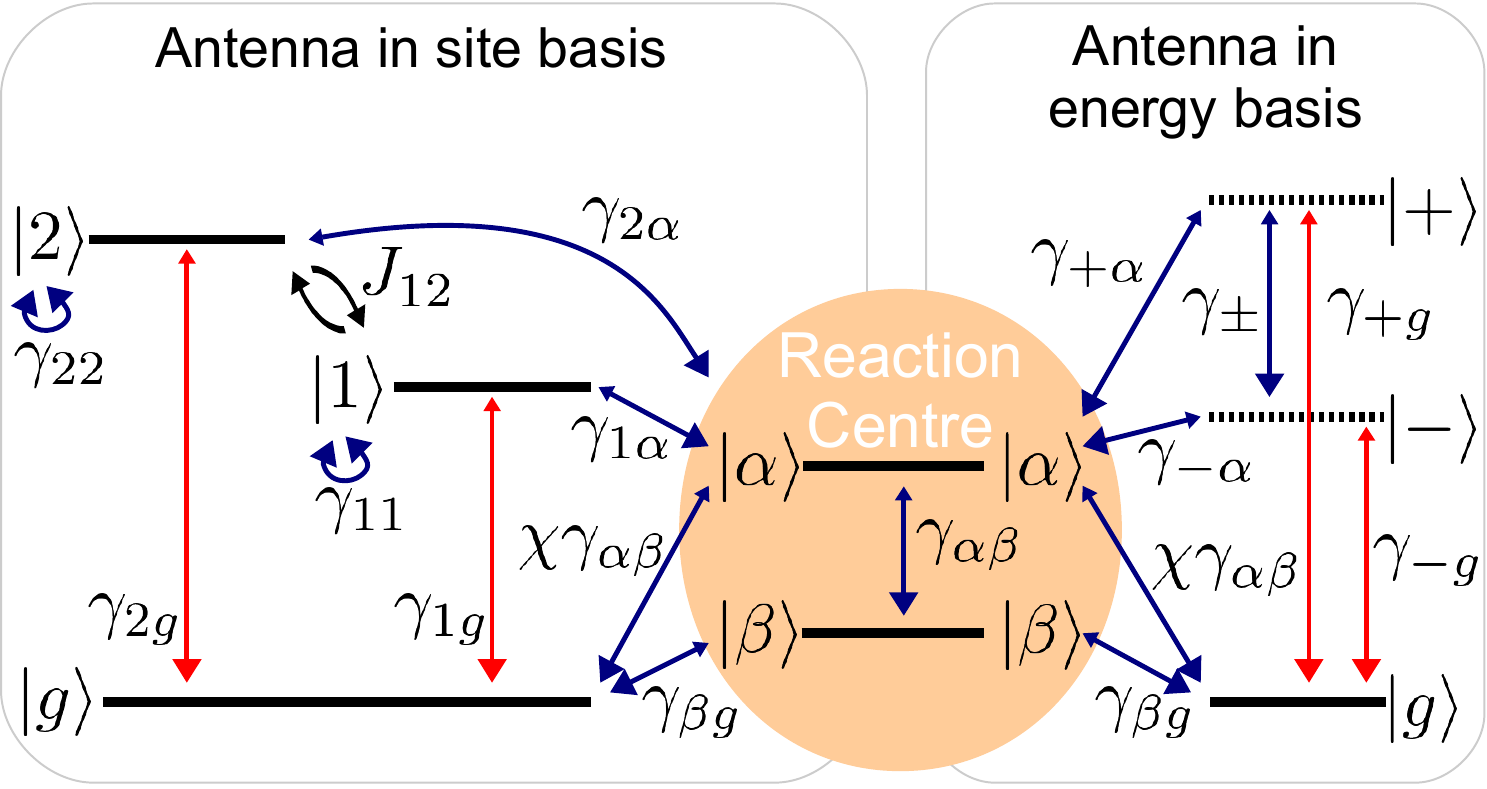}
\caption{\label{system schematics}
Level structure schematic of the three models. On the left we show the system in the site basis and on the right in the energy basis. The reaction centre is the same in both. We denote radiative transitions by red arrows, and non-radiative phonon-induced transitions by blue arrows.
}
\end{figure}

We denote the excited states in the molecules as $\ket{1}, \ket{2}$ with energies $\epsilon_1$, $\epsilon_2$, respectively, and the ground state as $\ket{g}$ with energy $\epsilon_g$, and $J_{12}$ is the dipolar coupling between the molecules. The Hamiltonian of the system is thus of dimension five and given by (see Fig.~\ref{system schematics})
\begin{align}
\Hamiltonian_s =& \epsilon_1 \ket{1}\bra{1} + \epsilon_2 \ket{2}\bra{2} + \epsilon_g \ket{g}\bra{g} 
\\ & + \frac{J_{12}}{2}\big(\ket{1}\bra{2}+\ket{2}\bra{1}\big) + \epsilon_\alpha \ket{\alpha}\bra{\alpha} + \epsilon_\beta \ket{\beta}\bra{\beta}
\nonumber \\
=& \epsilon_+ \ket{+}\bra{+} + \epsilon_- \ket{-}\bra{-} + \epsilon_g \ket{g}\bra{g}  \nonumber \\ 
&+\epsilon_\alpha \ket{\alpha}\bra{\alpha} + \epsilon_\beta \ket{\beta}\bra{\beta} ~.
\end{align}
In the second equation, $\ket{\pm}$ are the usual eigenstates diagonalising the subspace spanned by ${\ket{1}, \ket{2}}$.

In addition to the bare system we also have the solar photonic bath at $T_h = 6000$~K, which can induce spontaneous and stimulated transitions $\ket{1}\leftrightarrow \ket{g}$ and $\ket{2}\leftrightarrow \ket{g}$. Further, each molecule is embedded in its own local environment of vibrational modes, treated as infinite phonon baths at room temperature $T_c = 300$~K. A generic spin boson type interaction between excitonic states and phonon modes yields transitions between the energy eigenstates  $\ket{+}\leftrightarrow\ket{-}, \ket{+}\leftrightarrow\ket{\alpha},\ket{-}\leftrightarrow\ket{\alpha}, \ket{\beta}\leftrightarrow\ket{g}$~\cite{Gauger2010}. Further, we include the reaction center decay with rate $\gamma_{\alpha \beta}$, and some leakage between $\ket{\alpha}$ and $\ket{g}$ with rate $\chi \gamma_{\alpha \beta}$.
The interaction Hamiltonian is thus
\begin{align}
\Hamiltonian_I =& \hat I_{1g}+\hat I_{2g} +\hat I_{11}+\hat I_{22} +\hat I_{1\alpha} +\hat I_{2\alpha} + \hat I_{\beta g}
\end{align}
Here $\hat I_{ab} = \half(\ket{a}\bra{b}+\ket{b}\bra{a})\hat\mu_{ab}$, and $\hat\mu$ are 
operators of the coupling to the different environments:  $\hat\mu_{1 g},\hat\mu_{2 g}$ are dipole operators, and the rest are phonon operators~\footnote[1]{See Appendix \ref{Appendix B}}.

Applying a standard Born-Markov procedure \cite{BreuerPetruccione2007} we arrive at a set of Pauli master equations~\footnotemark{}:
\begin{gather}
\pd{}{t}\vec{P} = \mathcal{Q} \vec{P}.
\label{eq:rates}
\end{gather}
Here $\vec{P} = \{P_{+},P_{-},P_{\alpha},P_{\beta},P_{g}\}^\dagger$ is a vector of the populations in the diagonal basis of the system,
and $\mathcal{Q}$ is a matrix of the different rates, respecting detailed balance for photon and phonon baths independently. Resulting transition rates are depicted in Fig.~\ref{system schematics} and conservation of total population imposes the additional constraint $\sum_i P_i = 1$. We give the explicit entries of $\mathcal{Q}$ in Appendix \ref{Rate equation details}, and  also show that this rate equation approach is valid by direct comparison with the full Bloch-Redfield equations. 

Utilising the concept of a photochemical voltage \cite{Ross1967}, we attribute an effective current and voltage to the reaction center~\cite{Dorfman2012}:
\begin{gather}\label{Current Voltage def}
I = e \gamma_{\alpha \beta} P_\alpha, ~ V = \epsilon_\alpha - \epsilon_\beta - k_{\rm B} T_c \ln(P_\alpha / P_\beta) ~.
\end{gather}
Following Refs.~\onlinecite{Dorfman2012, Creatore2013,Killoran2015, Zhang2015, Stones2016} we now consider quantity $P = I V$ as a measure of the power generated by the system. Here $T_c$ is the (cold) phonon temperature, $k_{\rm B}$ the Boltzmann constant, and $e$ the electron charge.  In the absence of sunlight, the system thermalizes to the phonon bath temperature, and the voltage vanishes. Thus $V$ is a measure of the deviation from the thermal state with temperature $T_c$.
We are interested in the steady-state power output, which is found by setting the LHS of Eq.~(\ref{eq:rates}) to zero and solving the resulting simultaneous algebraic equations.

Our Hamiltonian is static, and so there is no cycle as there would be for conventional quantum heat engines~\cite{Niedenzu2015,Uzdin2015,Gelbwaser-Klimovsky2015}; our device rather relies on heat flowing through the reaction centre to produce work~\cite{Martinez2013,Kosloff2014,Levy2014}. Henceforth, we adopt the maximal achievable steady state power as our measure of efficiency: we treat the reaction centre as a black box optimizing its $\gamma_{\alpha \beta}$ to generate maximal power (keeping all other parameters fixed). For a representative example of the behaviour of $P$ and $I$ as a function of $V$ see Fig. \ref{IVPVplot}.

We now define the three specific models which we shall compare. These have contrasting molecular geometries and are depicted in Fig.~\ref{models comparison}. First, the {\it independent model} has two identical  light harvesting molecules that are not directly coupled to one another whilst each is independently coupled to the reaction centre. Specifically $\hat\mu_{1g} = \hat\mu_{2g}$; $\hat\mu_{1\alpha},\hat\mu_{2\alpha}$ represent the coupling to different phonon baths, and $J_{12} = \gamma_{+-} = 0$. Further, we let $\gamma_{+g} = \gamma_{-g}, \gamma_{+\alpha} = \gamma_{-\alpha}$, $\epsilon_+ = \epsilon_-$. This model does not exhibit dark state protection and serves as a benchmark for the other models.


Second, the {\it symmetric model} mirrors that described in Ref.~\onlinecite{Creatore2013} and consists of two identical, directly coupled molecules: $\epsilon_1 = \epsilon_2$, $\hat\mu_{1g} = \hat\mu_{2g}$, and  $J_{12}>0$. This arrangement leads to a dark and bright state, $\ket{+}$ and $\ket{-}$ respectively, with $\gamma_{-g}=0$ and $\gamma_{+-} = \frac{1}{4}(\gamma_{11}+\gamma_{22})$. The molecules couple to the reaction center in anti-phase $\hat\mu_{1\alpha} = -\hat\mu_{2\alpha}$~\cite{Creatore2013},
rendering $\gamma_{+\alpha}=0$. 
We discuss deviations from this idealized scenario in Appendix \ref{Appendix B}.

Finally, the {\it asymmetric model} is the main focus of our Letter. It comprises two non-identical molecules $\epsilon_1 < \epsilon_2$ with different dipole moments $\hat\mu_{1g} = z \hat\mu_{2g}$ where $z<1$ represents the asymmetry. The dark(er) $\ket{-}$ state has a larger overlap with molecule 1, which we imagine closer to the reaction center, and we assume molecule 2 only has negligibly small reaction center coupling ($\hat\mu_{2\alpha}=0$). We believe that this configuration should be easier to realise than the symmetric model, while allowing engineering of the energy gap $\epsilon_+-\epsilon_-$.
For flat spectral densities of the environments around the transition frequencies, we find this asymmetric model exhibits a fully dark state, provided that $J_{12}$, $z$, and $\epsilon_2-\epsilon_1$ satisfy the relation:
\begin{gather}\label{dark state criterion}
J_{12} = \frac{2z}{1-z^2} (\epsilon_2-\epsilon_1).
\end{gather}
Explicit rates for this system are given in Appendix \ref{Rate equation details}. Whether or not $\ket{-}$ is indeed fully dark, we can express the resulting total excitation rate through an angle $\Phi$:
\begin{align}
\gamma_{+g} =& (\gamma_{1g}+\gamma_{2g})\cos^2\Phi,\\
\gamma_{-g} =& (\gamma_{1g}+\gamma_{2g})\sin^2\Phi.
\end{align}
Thus $\tan^2\Phi = \gamma_{-g} / \gamma_{+g}$, and $\tan^2\Phi=0$ in the presence of a completely dark state. 

Several mechanisms may cause deviation from a fully dark state in both coupled models: First, different local environments would generally entail differing reorganization energy shifts and thus excitation energies. For example, the FMO complex consists of seven identical BChl units,  embedded in a protein scaffolding, resulting in on-site energies spanning a range of 25~meV~\cite{Adolphs2006}. 
Second, the two dipoles may be at an angle $\varphi$ instead of parallel~\citep{Creatore2013}, breaking the interference needed for a completely dark state. Third, the coherent coupling $J_{12}$ depends on both the distance between the two molecules, and the angle $\varphi$: $J_{12} = J_{12}^0 \cos\varphi$, where $J_{12}^0$ is the coupling with parallel dipoles. 
Taking all this into account we get in the general case, {\it i.e.} for all models,
\begin{align}\label{tan2thetaasym}
\tan^2&\Phi = \frac{\Omega_R(1+z^2) - (\epsilon_2-\epsilon_1)(1-z^2) - 2 z J_{12}\cos\varphi}{\Omega_R(1+z^2) + (\epsilon_2-\epsilon_1)(1-z^2) + 2 z J_{12}\cos\varphi},
\end{align}
with $\Omega_R = \sqrt{(\epsilon_2-\epsilon_1)^2+J_{12}^2}$ being the Rabi frequency of the bare system between sites 1 and 2. Further discussion about deviation from the fully dark state, and details on the coupling to the reaction centre are given Appendix \ref{Appendix B}.

The performance of the (a)symmetric relative to the independent models will be assessed by using our simulations to determine the ratio of the respective maximum powers, found by varying $\gamma_{\alpha \beta}$ in each case [see Eq.~(\ref{Current Voltage def})]. For a fair comparison, we keep $\gamma_{+g} + \gamma_{-g}$, $\epsilon_{-}$, and $\gamma_{1\alpha}$ equal across all three models. 

\section{Numerical results}\label{section numerical}

\begin{figure}
\centering
\includegraphics[width=\linewidth]{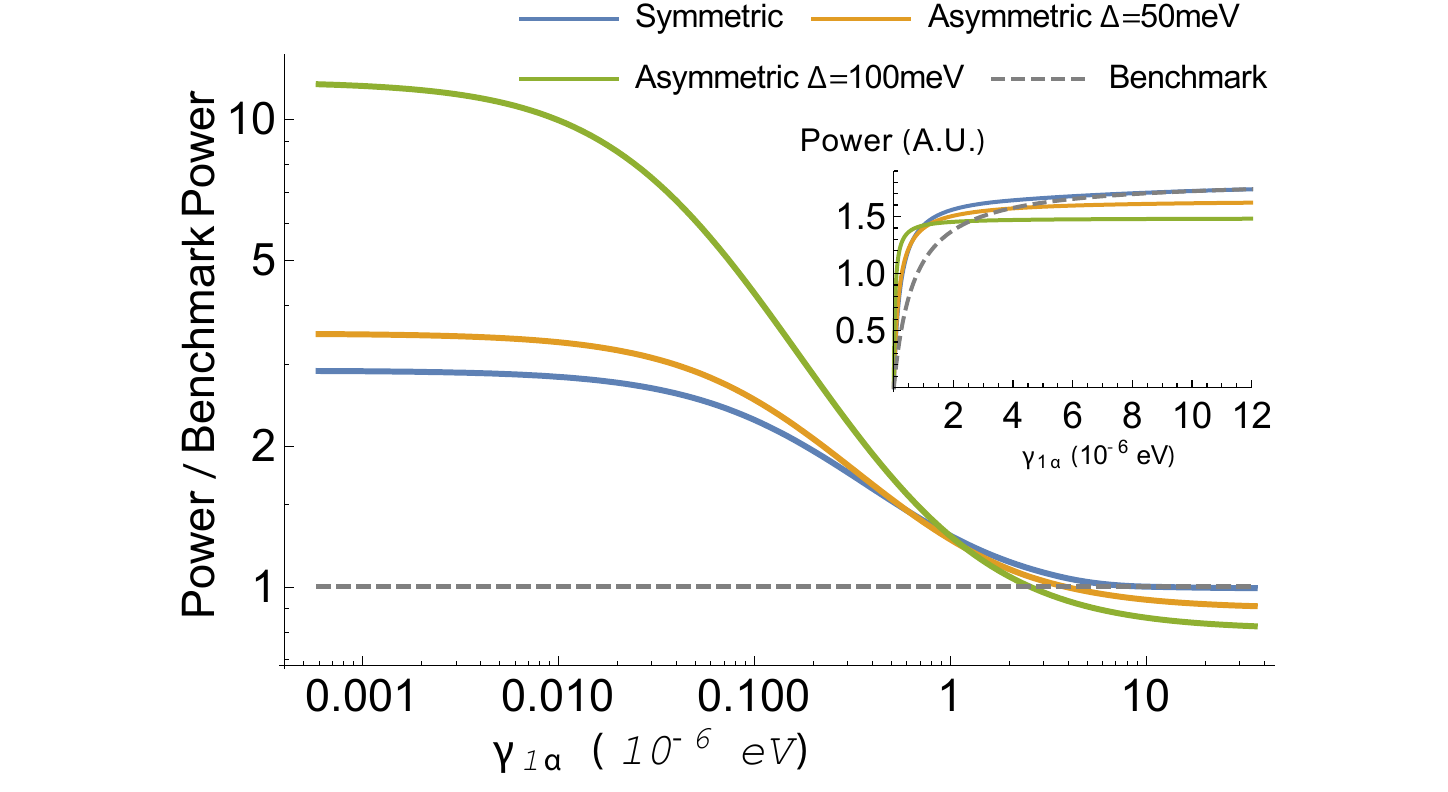}
\caption{\label{models relative enhancement}
Power enhancement over the benchmark achievable through a dark state, as a function of the trapping rate on a log-log scale.
The parameters were chosen to enable comparison with Ref.~\onlinecite{Creatore2013}:
$\gamma_{1g} + \gamma_{2g} = 1.24 \times 10^{-6}$~eV,
$\gamma_{11} = \gamma_{22} = 0.005$~eV,
$\gamma_{\beta g} = 0.0248$~eV,
$T_h = 6000$~K, $T_c = 300$~K,
$\epsilon_- = 2$~eV,
$\epsilon_\alpha = 1.8$~eV, $\epsilon_\beta = 0.2$~eV, $\chi = 0.2$.
Inset: the total power output of the systems in arbitrary units. 
}
\end{figure} 

Fig.~\ref{models relative enhancement} presents the enhancement achievable by dark-state protection. Here we have optimized $J_{12}$ as well as $\gamma_{\alpha \beta}$ with all other parameters fixed. We also constrain $J_{12}$ to below $30$~meV as an upper limit of realistic coupling strength. For the asymmetric model the dark state criterion, Eq.~\ref{dark state criterion}, informs an appropriate dipole asymmetry $z$ for a given value of $J_{12}$. 
Quantum enhancement is only possible when the transfer into the reaction center is relatively slow and constitutes a bottleneck in the cycle. In the limit of $\gamma_{1\alpha}\rightarrow 0$ an upper bound to the enhancement emerges. Within a reasonable parameter range this limit grows with increasing $J_{12}$ for the symmetric and with $\epsilon_2-\epsilon_1$ for the asymmetric case. As strong coupling is harder to realise than site energy mismatch, asymmetric dimers might more easily achieve high performance. We note that deeper into the slow transfer limit the potential enhancement factors can significantly exceed the values of up to 50\% reported by Refs.~\onlinecite{Dorfman2012, Creatore2013, Zhang2015}. By contrast, for fast transfer rates dark-state protection offers no advantage: absorbing photons at an energy higher than the extraction energy ({\it i.e.} at reduced thermal photon occupancy), combined with $\gamma_{2\alpha} = 0$, is now detrimental.

\begin{figure}
\centering
\includegraphics[width=\linewidth]{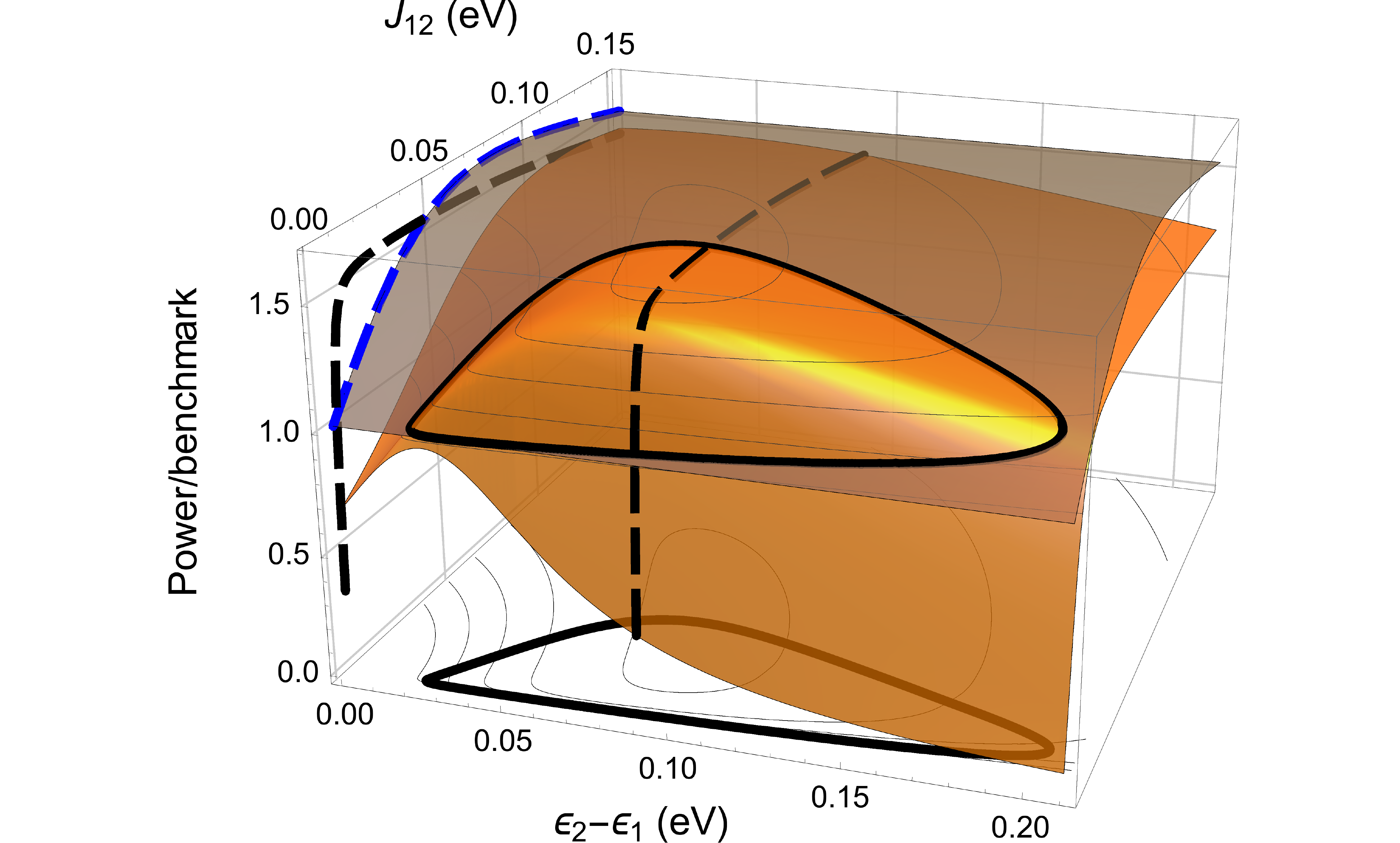}
\caption{\label{3dplot}
Orange surface: relative power enhancement of the asymmetric model, as a function of the energy difference $\epsilon_2-\epsilon_1$ and coupling $J_{12}$. Gray surface:  relative power enhancement of the symmetric model (independent of $\epsilon_2 - \epsilon_1$). Black dashed line: asymmetric model enhancement for a fixed $\epsilon_2-\epsilon_1 = 90$~meV. The dashed line is projected onto the $\epsilon_2-\epsilon_1 = 0$ plane, for comparison with the symmetric equivalent (dashed blue). The black thick line is the contour where the symmetric and asymmetric power ratios are equal.
Parameters are as in Fig.~\ref{models relative enhancement} and with $\gamma_{1\alpha} = 6\times  10^{-7}$~eV.
}
\end{figure}

Fig.~\ref{3dplot} shows the relative power enhancements of the asymmetric model as a function of the energy difference $\epsilon_1-\epsilon_2$ and of the coupling $J_{12}$. We also plot the equivalent enhancement given by the symmetric model, and show there is a parameter regime, with boundaries marked by a black line, for which the asymmetric model outperforms the symmetric one. 
The asymmetric model  displays a peak power enhancement at finite coupling and energy difference. This happens for two reasons: First, in the regime $\epsilon_+-\epsilon_- \lesssim k_{\rm B}T$, the rate $\ket{-}\rightarrow\ket{+}$ is non-negligible, and so the dark state is not protected. Second, if $\epsilon_+-\epsilon_- \gg J_{12}$, the rate $\ket{+}\rightarrow\ket{-}$ becomes negligible, and the dark state is rarely populated. Note that in the limit $J_{12}\rightarrow 0$, the asymmetric model gives a smaller power than the independent benchmark. This is because we set $\gamma_{2\alpha}=0$ for the asymmetric model, effectively making it a single antenna setup benchmarked against two antennae. We examine further realistic imperfections, including the presence of additional dephasing mechanisms,~in Appendix \ref{Appendix B}.

\section{Candidate molecules for realizing the asymmetric model}\label{section rafa}

To assess the feasibility of generating suitable asymmetric molecules for power enhancement, we use a library containing quantum-chemically predicted properties of organic light-emitting diode molecules~\cite{GomezBombarelli2016} to  identify systems that minimize $\tan^2 \Phi$ in Eq.~(\ref{tan2thetaasym}). Appendix \ref{Appendix A} provides a full account of how quantum chemical calculations lead to promising molecular dimer candidates through a rigorous multistage process. Our donor candidates feature strongly allowed optical transitions ($\mu$ of 3.5 atomic units) and site energies between 3.5 to 2.5 eV to optimally absorb sunlight. As required by the model, acceptable acceptor compounds must have $\epsilon_1 < \epsilon_2$ and possess lower transition dipole moments with $z \approx 0.2$ to deliver $\tan^2\Phi \lesssim 0.05$. 
For simplicity, we assume fully aligned transition dipole moments  and center-to-center distances between donor and acceptor moieties of 1~nm (approximately corresponding to the size of a small aromatic bridging group), resulting in an inter-site coupling of up to 15~meV. Importantly, we analyse the properties of our dimers for both ground and excited state equilibrium geometry to identify systems whose relevant properties are robust to vibrational relaxation effects accompanying optical absorption and emission.

The predicted properties of a selection of molecular pairs are reported in Table \ref{pairs table} of Appendix \ref{Appendix A} and an illustration of some molecules is shown in Fig.~\ref{fig:moldensity}. These examples provide evidence that the chemical regime required for dark-state protection is readily available in ordinary molecular systems. A full implementation of the proposed model would also require an additional molecular system to act as a trap, as well as control over orientation and distance between donor and acceptor.  Whereas the chemical synthesis of such a complex structure is challenging, our results show that matching fundamental components for such a system is entirely feasible.

\section{Summary and Conclusions}\label{section summary}

In conclusion, we have presented a general model of light absorption by an asymmetric pair of coupled chromophores, finding that it can outperform both the symmetric dimer and a pair of independent molecules in realistic parameter regimes of operation for a solar cell device. Not relying on identically matched coupled chromophores, this approach is more robust to deviations from the delicate conditions required by its symmetric counterpart. Moreover, we have shown that an abundance of real pairs of molecules have the required asymmetric properties, and indeed, such asymmetry is an integral part of natural photosynthetic systems. 

The reason our asymmetric model works so well is that it enables arbitrarily large energy gaps between the bright and dark states, thus preventing phonon-assisted promotion from the dark to the bright state. In the regime where excitations are rare and the transfer into the reaction center is very slow, this translates into better protection of the excitations, thus increasing the overall efficiency of the device.

\begin{acknowledgments}
We thank Alex Chin, Ahsan Nazir, Simon Benjamin, Al\'an Aspuru-Guzik, and Jorge Aguilera-Iparraguirre for stimulating discussions. This work was supported by the Leverhulme Trust (RPG-080).
EMG is supported by the Royal Society of Edinburgh / Scottish Government. 
RGB thanks Samsung Advanced Institute of Technology for funding.
AF thanks the Anglo-Israeli association and the Anglo-Jewish association for funding.
\end{acknowledgments}

\appendix
\section{Molecular dimers candidates}\label{Appendix A}

In the following we discuss a possible roadmap towards finding candidate dimer systems based on pairs of real molecules, which are predicted to display the desired dark-state protection effect.  Our pool of molecules consists of a database listing 500,000 viable organic dye molecules. Their site energies and transition dipole moments were calculated at the TD-DFT B3LYP/6-31G(d) level of theory on molecular geometries optimized at the B3LYP/6-31G(d) level.\cite{GomezBombarelli2016}

\begin{figure}
\centering
\includegraphics[width=0.49\textwidth]{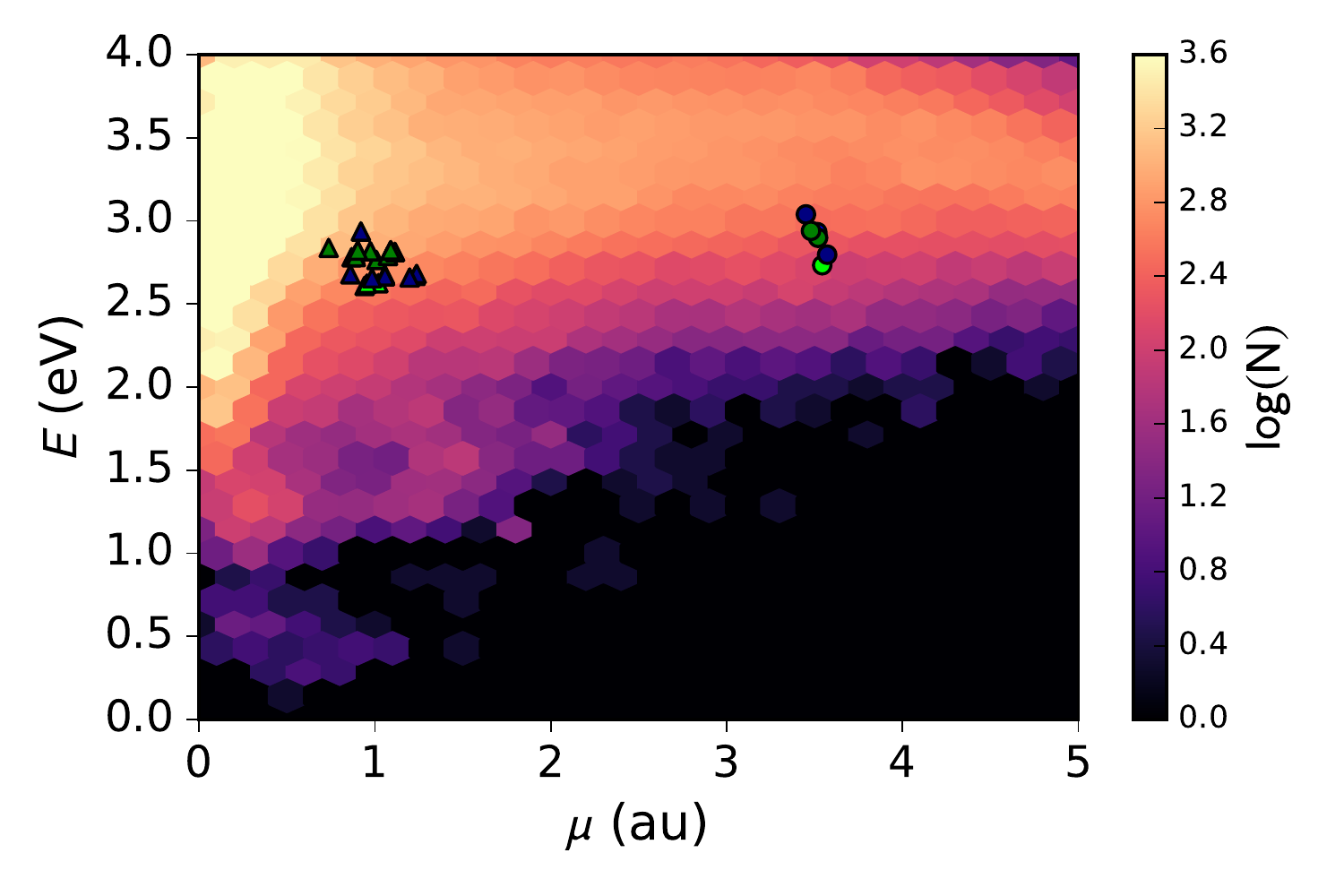}
\caption{\label{fig:moldensity}
Density of calculated dye molecules for a given transition dipole moment and site energy at the TD-DFT B3LYP/6-31G(d)//B3LYP/6-31G(d) level of theory. Lead donor and acceptor pairs are marked as circles and triangles of matching color.}
\end{figure}

As a starting point in our endeavour to identify potential dimer candidates, we survey the relevant optical properties of individual molecules. Figure.~\ref{fig:moldensity} shows the distribution of absorption energy vs optical dipole strength across all molecules in the database.

\begin{figure*}
\centering
\includegraphics[width=0.49\textwidth]{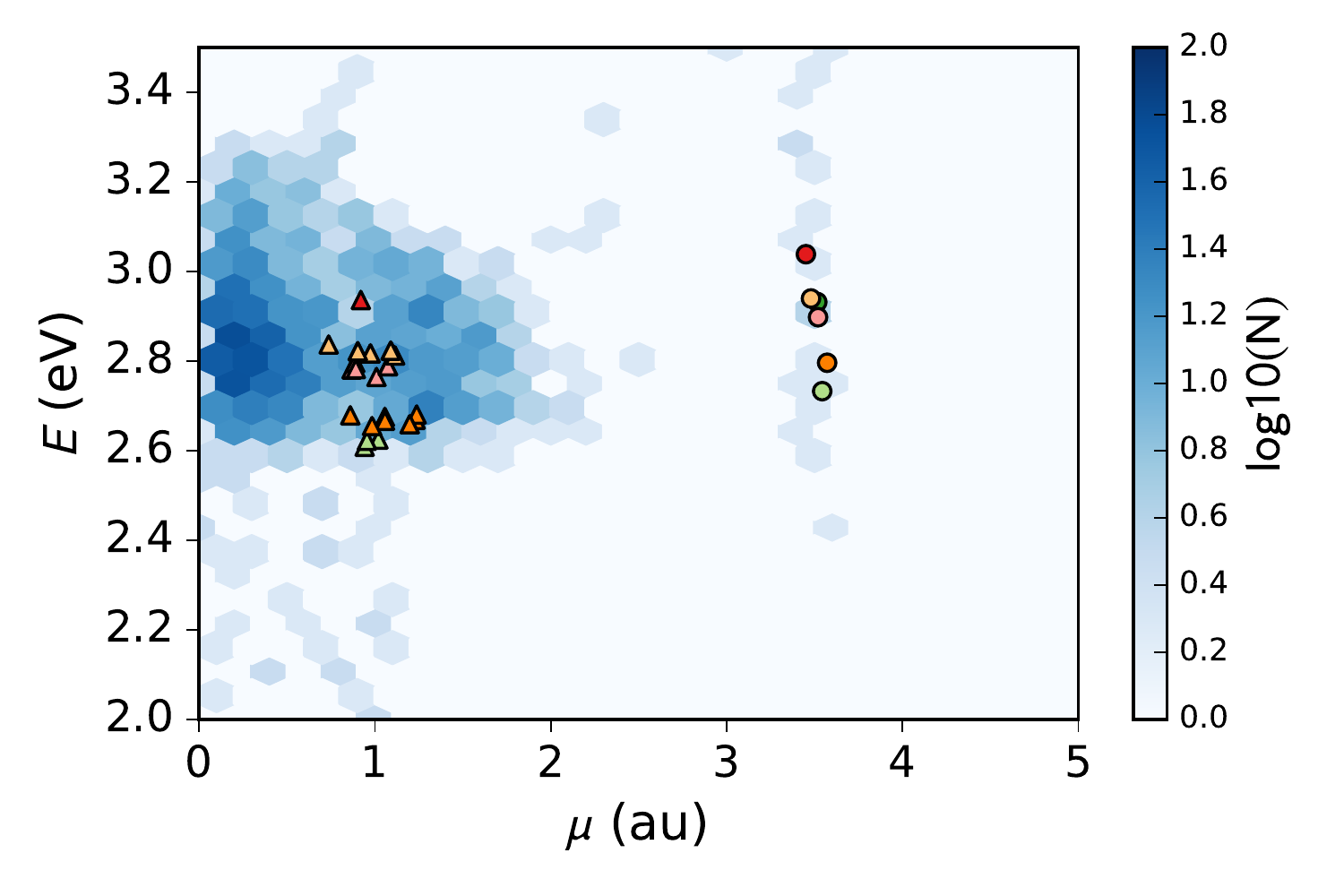}
\includegraphics[width=0.49\textwidth]{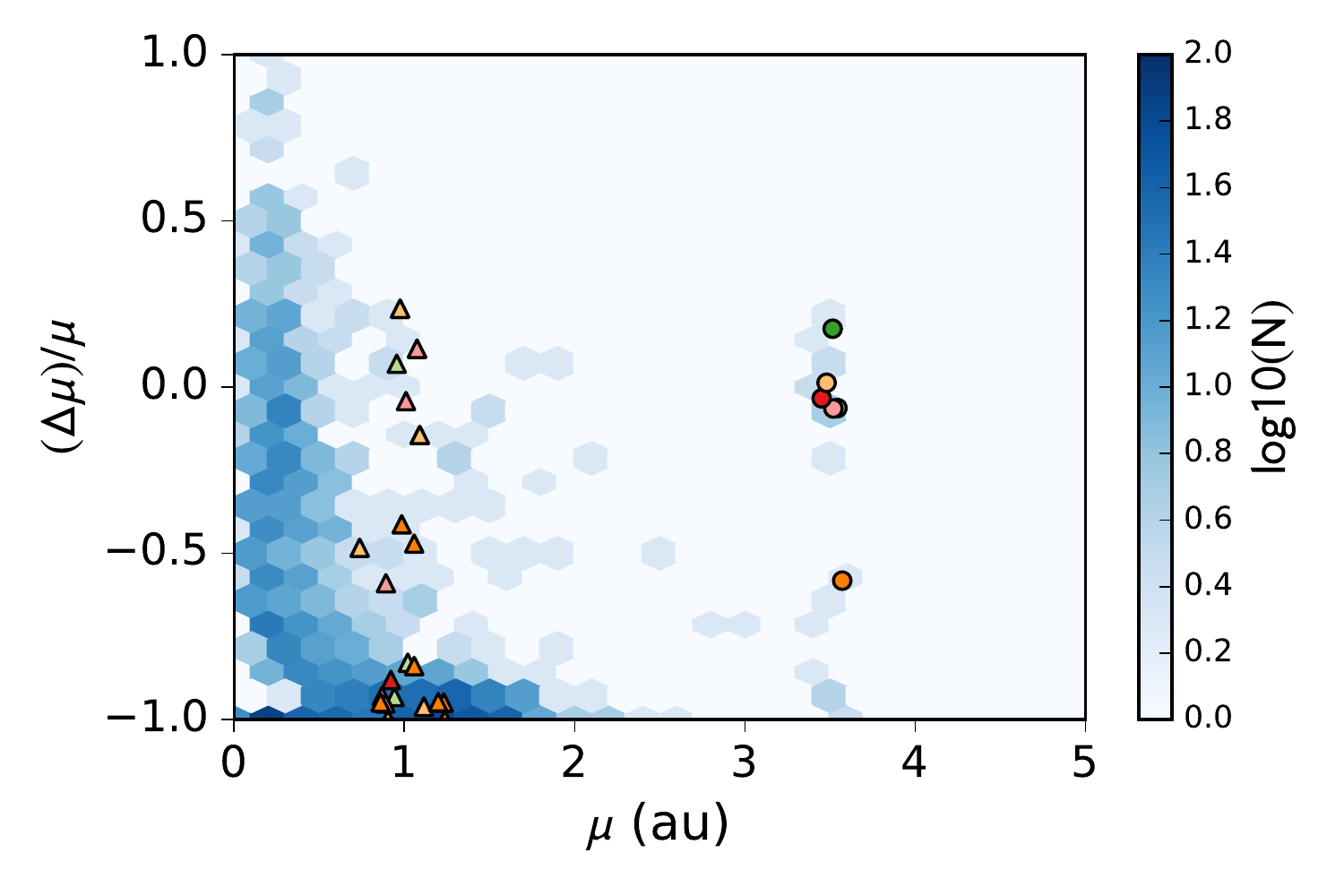}
\includegraphics[width=0.49\textwidth]{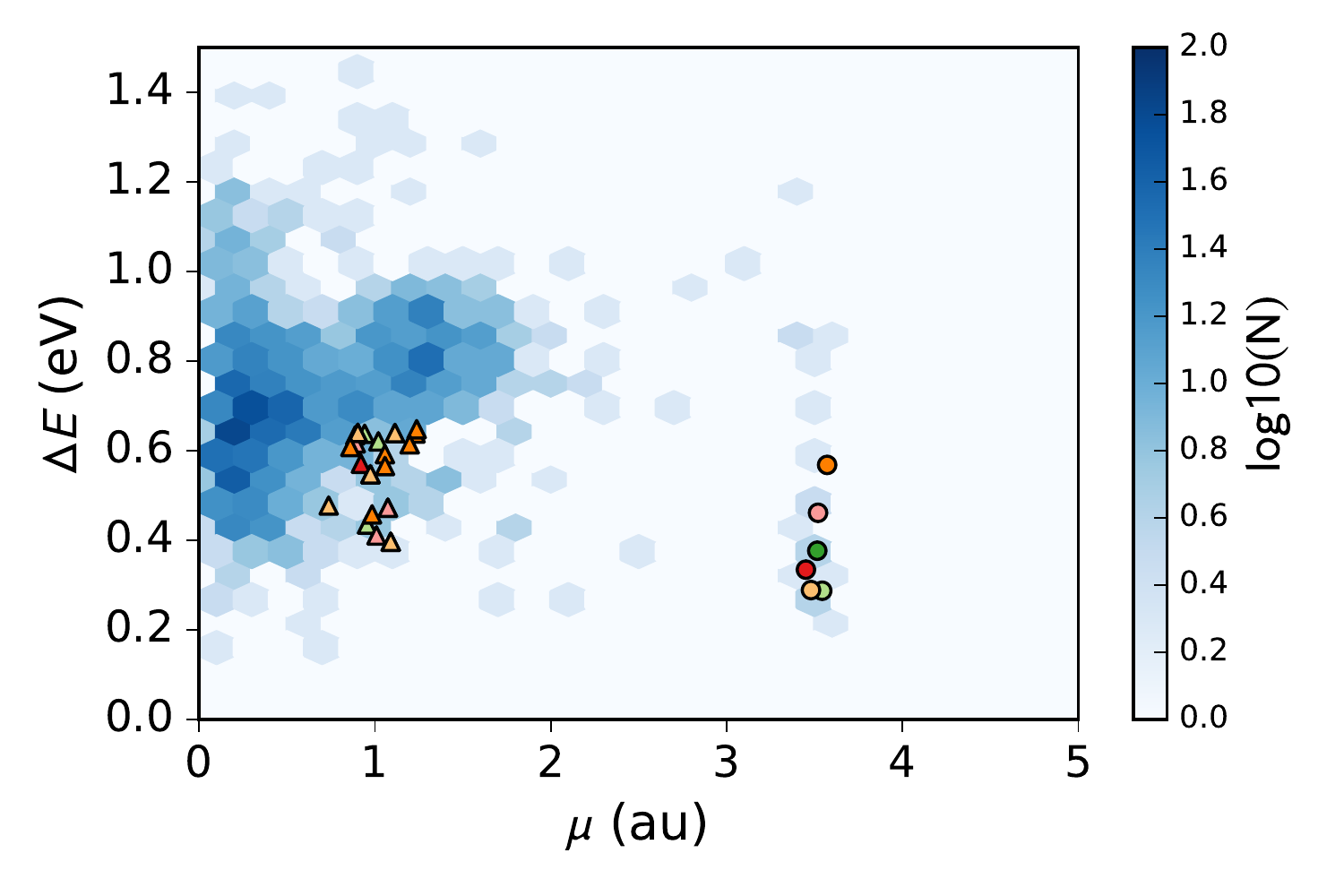}
\includegraphics[width=0.49\textwidth]{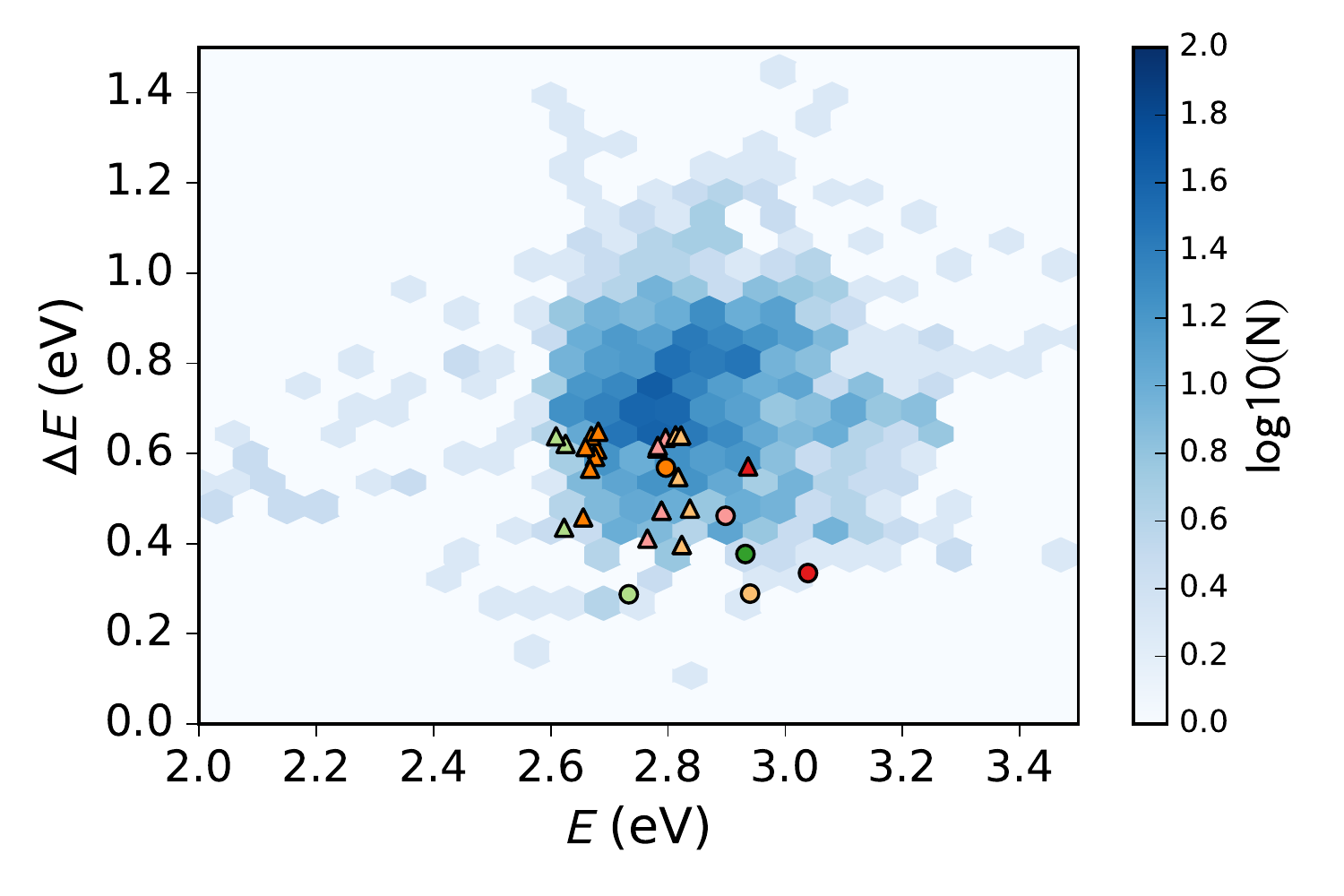}
\caption{\label{fig:relaxed_moldensity}
Plots of key optical properites of candidate molecules. Upper left: absorption energy against dipole strength; upper right: relative difference in dipole strength between absorption and emission, as a function of dipole strength; lower panels: Stokes shift as a function of dipole strength (left) and transition energy (right). 
Lead donor and acceptors pairs are marked as circles and triangles of the same color. 
}
\end{figure*}

Recognising that vibronic effects are usually important in molecular systems, we have calculated the effect of geometric relaxation on a selection of 2,000 molecules by optimizing the molecular geometries of their lowest excited state at the TD-DFT B3LYP/6-31G(d) level of theory. In Fig.~\ref{fig:relaxed_moldensity} we survey the relevant optical properties of this reduced sample. We plot the absorption energy as a function of optical dipole strength (upper left panel), the relative difference in dipole strength between absorption and emission (upper right panel), and the Stokes shift as a function of dipole strength and transition energy, respectively, (lower panels).

Of relevance to our scheme, we note there is an abundance of molecules with absorption energies in the 2.5-3.5 eV bracket --- {\it i.e.},~near the power maximum in the spectrum of sunlight --- and that these molecules possess a varying degree of dipole strength. There are also a sizeable number of candidates with relatively low Stokes shift, mainly appearing in two clusters with comparatively strong and weak optical dipoles. Finally, candidates exist for which there is only a small difference in the dipole moments of absorption and emission, consistent with dipoles that have little dependence on vibronic coupling, suggesting that the nature of the relevant states will not change significantly with geometrical relaxation.

\begin{figure*}
\centering
\includegraphics[width=0.45\textwidth]{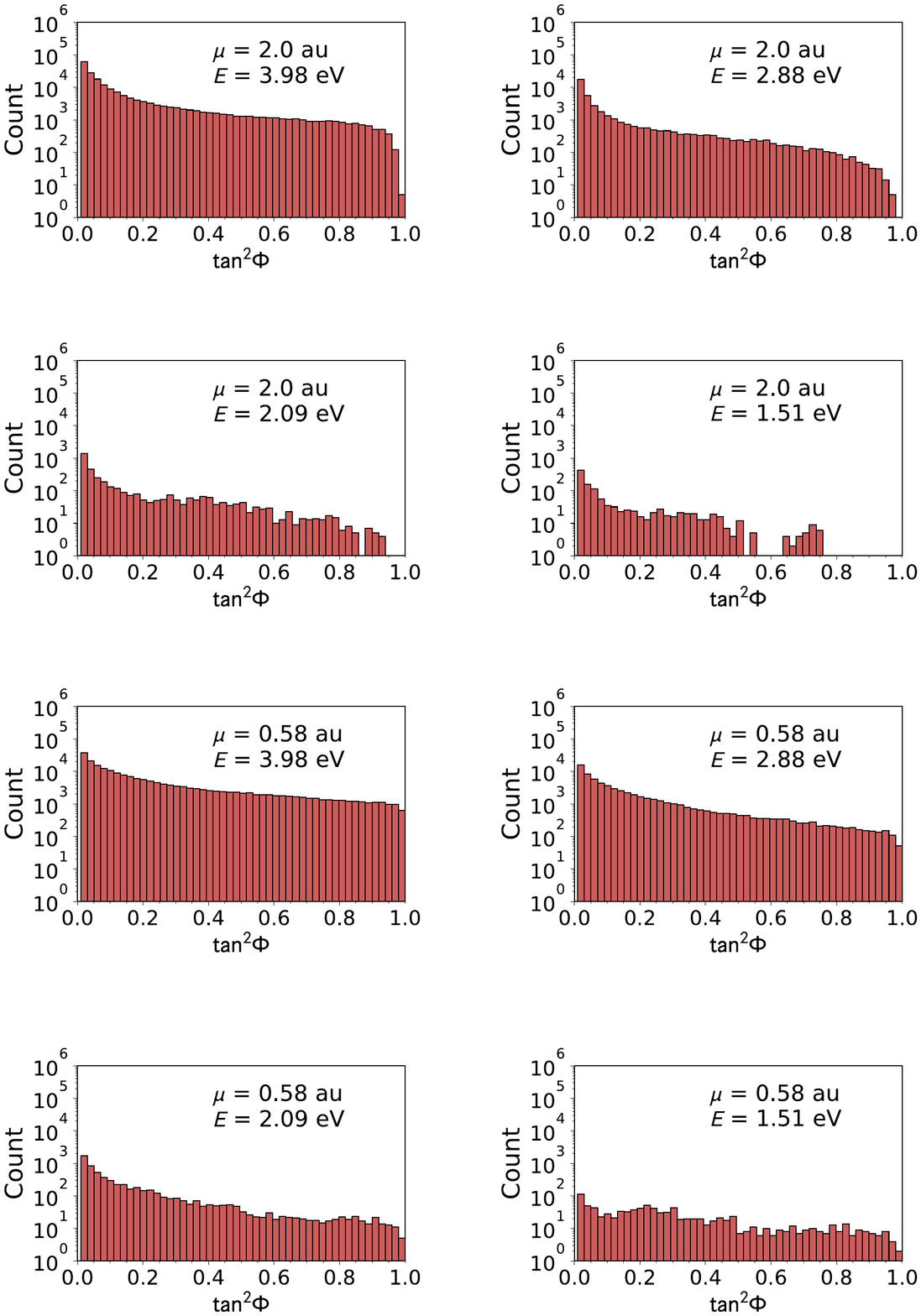}
\includegraphics[width=0.45\textwidth]{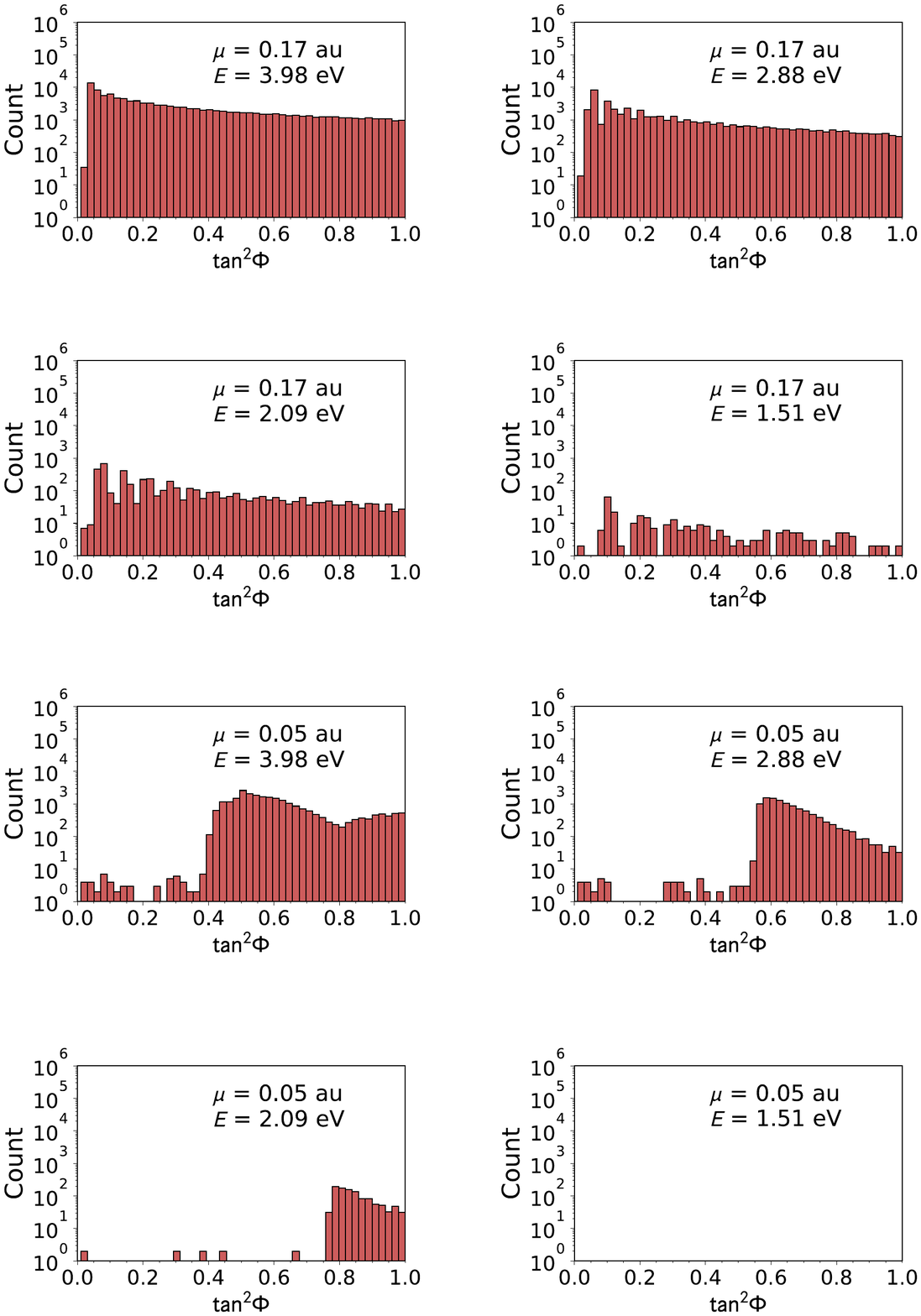}
\includegraphics[width=0.45\textwidth]{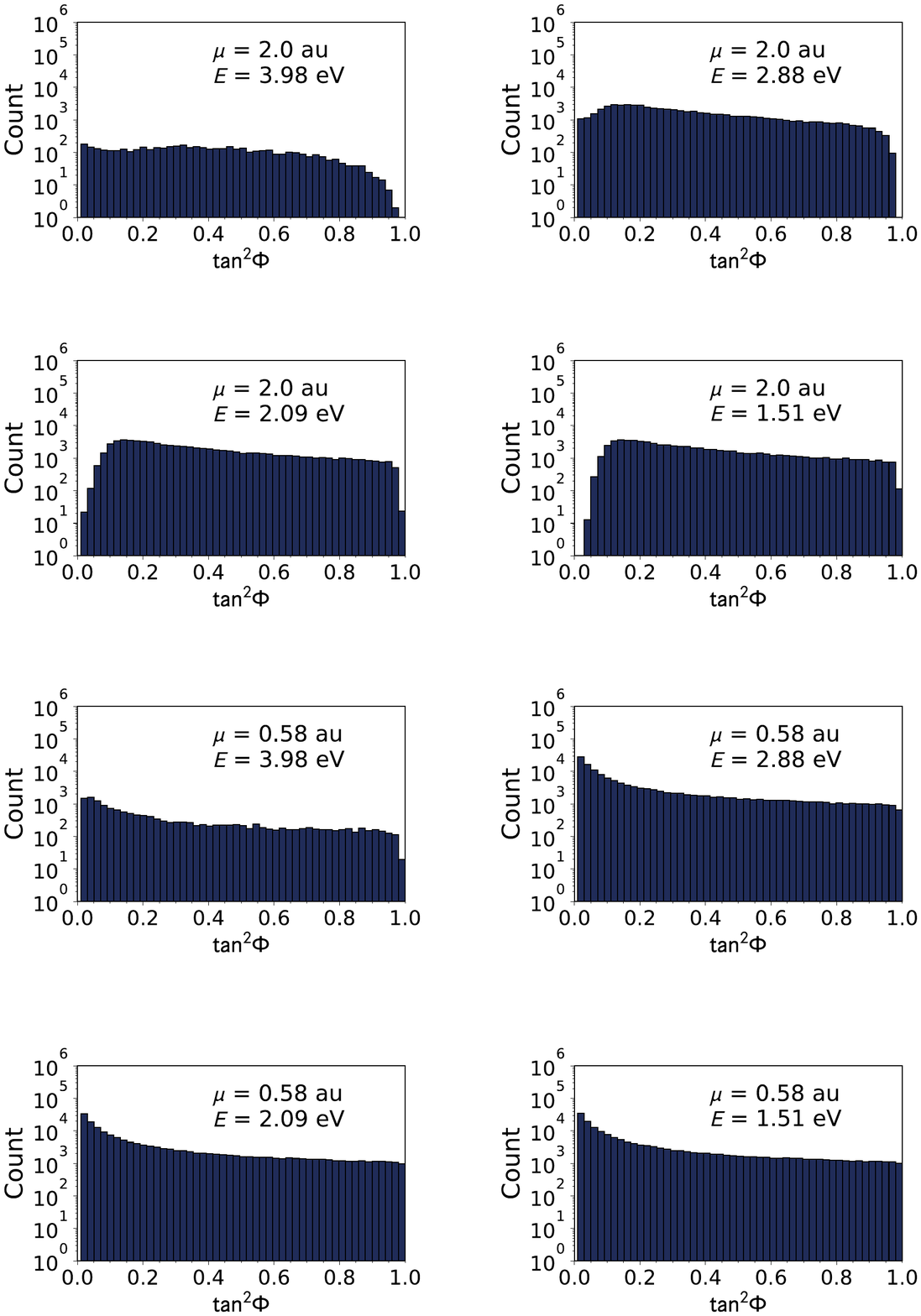}
\includegraphics[width=0.45\textwidth]{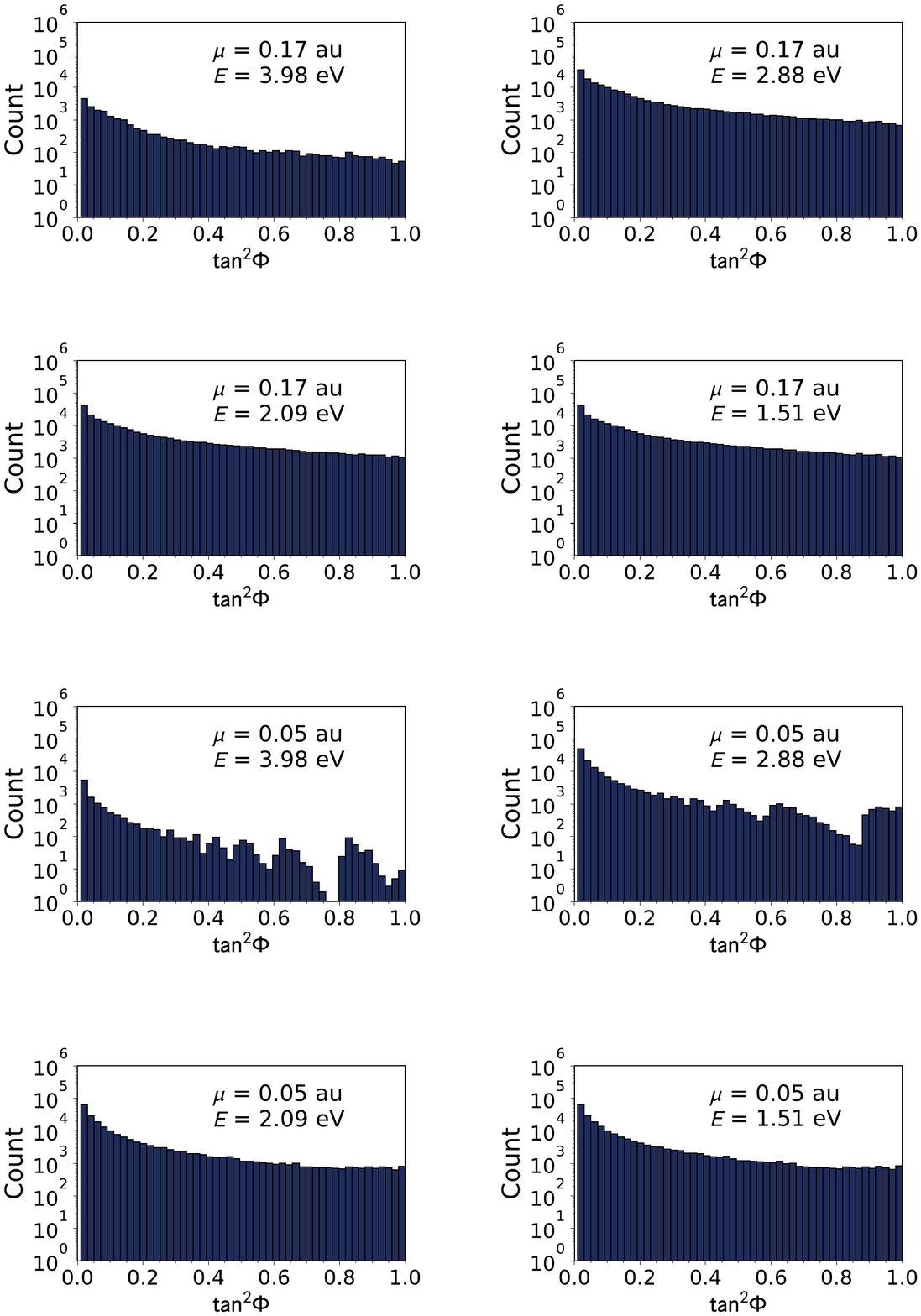}
\caption{\label{fig:donors_for_acceptor}
Top: Abundance of potential acceptor molecules in the molecular database, given a donor with transition dipole ($\mu$) and site energy ($E$), as a function of $\tan^2 \Phi$.
Bottom: Abundance of potential donor molecules in the molecular database, given an acceptor with transition dipole ($\mu$) and site energy ($E$), as a function of $\tan^2 \Phi$.
}
\end{figure*}

We now attempt to pair up molecules according to whether they show promise for displaying dark-state behaviour.  For this purpose the donor and acceptor are assumed to be coupled via point dipole coupling of their lowest local excited states. For computational ease of screening the still large set of candidates, we base our estimate of the F\"orster coupling on the oscillator strength and transition energy of the absorption transition (S0). The histograms displayed in Fig.~\ref{fig:donors_for_acceptor} report the density of suitable donor (or acceptor) partners in the database for a selection of 16 given acceptors (or donors) with a wide range of properties. As shown, a very large number of potential partner molecules with a high level of predicted dark state protection ($\tan^2 \Phi \lesssim 0.05$) exists in the majority of all cases considered.

To take a further step, we report a selection of 26 candidate chromophore pairs which are expected to be suitable for dark-state protection according to the predicted properties on the ground state equilibrium geometry (all rows in Table~\ref{pairs table}). These 26 candidates were obtained as combinations of the 6 donor and 20 acceptor molecules highlighted in Figs.~\ref{fig:moldensity} and \ref{fig:relaxed_moldensity}. For these candidates we also calculate the emission properties, based on the excited state equilibrium geometry (results again in Table~\ref{pairs table}). This shows that not all pairs can be expected to be fully characterised by our simpler initial approach: Some molecules, particularly acceptors, have high vibronic coupling and experience large changes in both site energy and transition dipole upon excitation. 

To screen further, we therefore select the six pairs with the overall most favourable values for $\tan^2 \Phi$ across both excited and ground geometries. These highly promising candidate pairs are printed in rows A-F of Table~\ref{pairs table}, combining a pronounced dark-state both from the S0 and the S1 transition. No large-scale vibronic effects, such as state reordering are observed for the intermediate geometries for these pairs. We believe these dimer candidates would be excellent candidates for full experimental and/or \textit{ab-initio} computational characterisation, as in intermediate goal on route to a full experimental implementation of our proposed scheme. 

We note that throughout our process of narrowing down the field, we have at several stages had to reduce the pool of candidates by random selection, so our final list of six candidates will likely only be a tiny fraction of interesting dimer combinations existing in the full space of the database. Further, fluorescent dyes, such as cyanine or BODIPY derivatives were not explored in this particular dataset, but are also expected to produce species with large transition dipole moment at arbitrary absorption energies. 

\begin{table*}
\begin{center} 
    \begin{tabular}{ | c |  c |  c |  c |  c |  c |  c |  c |  c |  c |  c |  c |  c |  c |  c |  c |  c |  c | c | }
    \hline
\# & $E_2^g$ & $\mu_2^g$ & $E_2^e$ & $\mu_2^e$ & $E_1^g$ & $\mu_1^g$ & $E_1^s$ & $\mu_1^e$ & $z_g$ & $z_e$ & $J_g$ & $J_e$ & $\tan^2\Phi_g$ & $\tan^2\Phi_e$ & Q\\ \hline
1 & 2.73 & 3.54 & 2.45 & 3.32 & 2.61 & 0.94 & 1.97 & 0.06 & 0.27 & 0.02 & 0.013 & 0.012 & 0.044 & 0.000 & 1.41\\ \hline
2 & 2.73 & 3.54 & 2.45 & 3.32 & 2.63 & 1.02 & 2.00 & 0.17 & 0.29 & 0.05 & 0.015 & 0.013 & 0.047 & 0.001 & 1.45\\ \hline
3 & 2.93 & 3.52 & 2.56 & 4.13 & 2.82 & 0.90 & 2.18 & 0.00 & 0.26 & 0.00 & 0.013 & 0.018 & 0.038 & 0.001 & 1.45\\ \hline
4 & 2.93 & 3.52 & 2.56 & 4.13 & 2.82 & 1.09 & 2.43 & 0.93 & 0.31 & 0.23 & 0.015 & 0.021 & 0.055 & 0.020 & 1.43\\ \hline
5 & 2.93 & 3.52 & 2.56 & 4.13 & 2.81 & 1.11 & 2.17 & 0.04 & 0.32 & 0.01 & 0.016 & 0.022 & 0.060 & 0.000 & 1.40\\ \hline
6 & 2.90 & 3.52 & 2.44 & 3.29 & 2.80 & 0.89 & 2.16 & 0.04 & 0.25 & 0.01 & 0.013 & 0.011 & 0.035 & 0.000 & 1.47\\ \hline
7 & 2.90 & 3.52 & 2.44 & 3.29 & 2.78 & 0.87 & 2.17 & 0.06 & 0.25 & 0.02 & 0.012 & 0.011 & 0.037 & 0.000 & 1.44\\ \hline
8 & 2.90 & 3.52 & 2.44 & 3.29 & 2.78 & 0.89 & 2.17 & 0.37 & 0.25 & 0.11 & 0.013 & 0.011 & 0.038 & 0.008 & 1.44\\ \hline
9 & 3.04 & 3.45 & 2.70 & 3.33 & 2.94 & 0.92 & 2.37 & 0.11 & 0.27 & 0.03 & 0.013 & 0.012 & 0.040 & 0.000 & 1.46\\ \hline
10 & 2.94 & 3.48 & 2.65 & 3.52 & 2.84 & 0.74 & 2.36 & 0.38 & 0.21 & 0.11 & 0.010 & 0.011 & 0.026 & 0.008 & 1.47\\ \hline
11 & 2.94 & 3.48 & 2.65 & 3.52 & 2.82 & 0.90 & 2.18 & 0.00 & 0.26 & 0.00 & 0.013 & 0.013 & 0.041 & 0.000 & 1.43\\ \hline
12 & 2.94 & 3.48 & 2.65 & 3.52 & 2.82 & 0.97 & 2.27 & 1.20 & 0.28 & 0.34 & 0.014 & 0.014 & 0.049 & 0.103 & 1.41\\ \hline
13 & 2.94 & 3.48 & 2.65 & 3.52 & 2.81 & 1.11 & 2.17 & 0.04 & 0.32 & 0.01 & 0.016 & 0.016 & 0.065 & 0.000 & 1.38\\ \hline
14 & 2.80 & 3.57 & 2.23 & 1.49 & 2.68 & 0.86 & 2.07 & 0.04 & 0.24 & 0.03 & 0.012 & 0.002 & 0.035 & 0.001 & 1.44\\ \hline
15 & 2.80 & 3.57 & 2.23 & 1.49 & 2.66 & 0.98 & 2.20 & 0.58 & 0.28 & 0.39 & 0.014 & 0.002 & 0.049 & 0.117 & 1.36\\ \hline
16 & 2.80 & 3.57 & 2.23 & 1.49 & 2.68 & 1.06 & 2.08 & 0.56 & 0.30 & 0.38 & 0.015 & 0.003 & 0.052 & 0.133 & 1.41\\ \hline
17 & 2.80 & 3.57 & 2.23 & 1.49 & 2.67 & 1.06 & 2.10 & 0.17 & 0.30 & 0.11 & 0.015 & 0.003 & 0.054 & 0.011 & 1.39\\ \hline
18 & 2.80 & 3.57 & 2.23 & 1.49 & 2.68 & 1.24 & 2.03 & 0.00 & 0.35 & 0.00 & 0.018 & 0.003 & 0.069 & 0.000 & 1.40\\ \hline
19 & 2.80 & 3.57 & 2.23 & 1.49 & 2.66 & 1.20 & 2.05 & 0.06 & 0.34 & 0.04 & 0.017 & 0.003 & 0.071 & 0.001 & 1.35\\ \hline
20 & 2.80 & 3.57 & 2.23 & 1.49 & 2.67 & 1.23 & 2.03 & 0.06 & 0.34 & 0.04 & 0.018 & 0.003 & 0.072 & 0.001 & 1.37\\ \hline
A & 2.73 & 3.54 & 2.45 & 3.32 & 2.62 & 0.95 & 2.19 & 1.02 & 0.27 & 0.31 & 0.014 & 0.012 & 0.042 & 0.080 & 1.45\\ \hline
B & 2.93 & 3.52 & 2.56 & 4.13 & 2.82 & 0.97 & 2.27 & 1.20 & 0.28 & 0.29 & 0.014 & 0.019 & 0.046 & 0.065 & 1.43\\ \hline
C & 2.93 & 3.52 & 2.56 & 4.13 & 2.79 & 1.07 & 2.32 & 1.20 & 0.31 & 0.29 & 0.015 & 0.021 & 0.062 & 0.059 & 1.34\\ \hline
D & 2.90 & 3.52 & 2.44 & 3.29 & 2.76 & 1.01 & 2.35 & 0.97 & 0.29 & 0.29 & 0.014 & 0.013 & 0.053 & 0.046 & 1.38\\ \hline
E & 2.90 & 3.52 & 2.44 & 3.29 & 2.79 & 1.07 & 2.32 & 1.20 & 0.31 & 0.36 & 0.015 & 0.013 & 0.053 & 0.092 & 1.43\\ \hline
F & 2.94 & 3.48 & 2.65 & 3.52 & 2.82 & 1.09 & 2.43 & 0.93 & 0.31 & 0.26 & 0.015 & 0.016 & 0.059 & 0.052 & 1.41\\ \hline
    \end{tabular}
\end{center}
\caption{\label{pairs table}
List of 26 candidate pairs which exhibit a relatively good dark state, with their relative enhancement to the benchmark. The symbols are as defined in the main text with the number index 2 and 1 labelling donor and acceptor, respectively.  The super- and subscripts $e$ and $g$ denote whether the relevant property was obtained from the relaxed excited or ground state geometry, i.e.~from the S1 or the S0 transition. The final six pairs A-F  have only small differences in $\tan^2\Phi_g$ and $\tan^2\Phi_e$ for the ground and excited state geometries, suggesting the nature of the relevant states will be robust to vibronic relaxation. Q = power / benchmark power for the ground state geometries.
}\end{table*}

\begin{figure}
\centering
\includegraphics[scale=0.45]{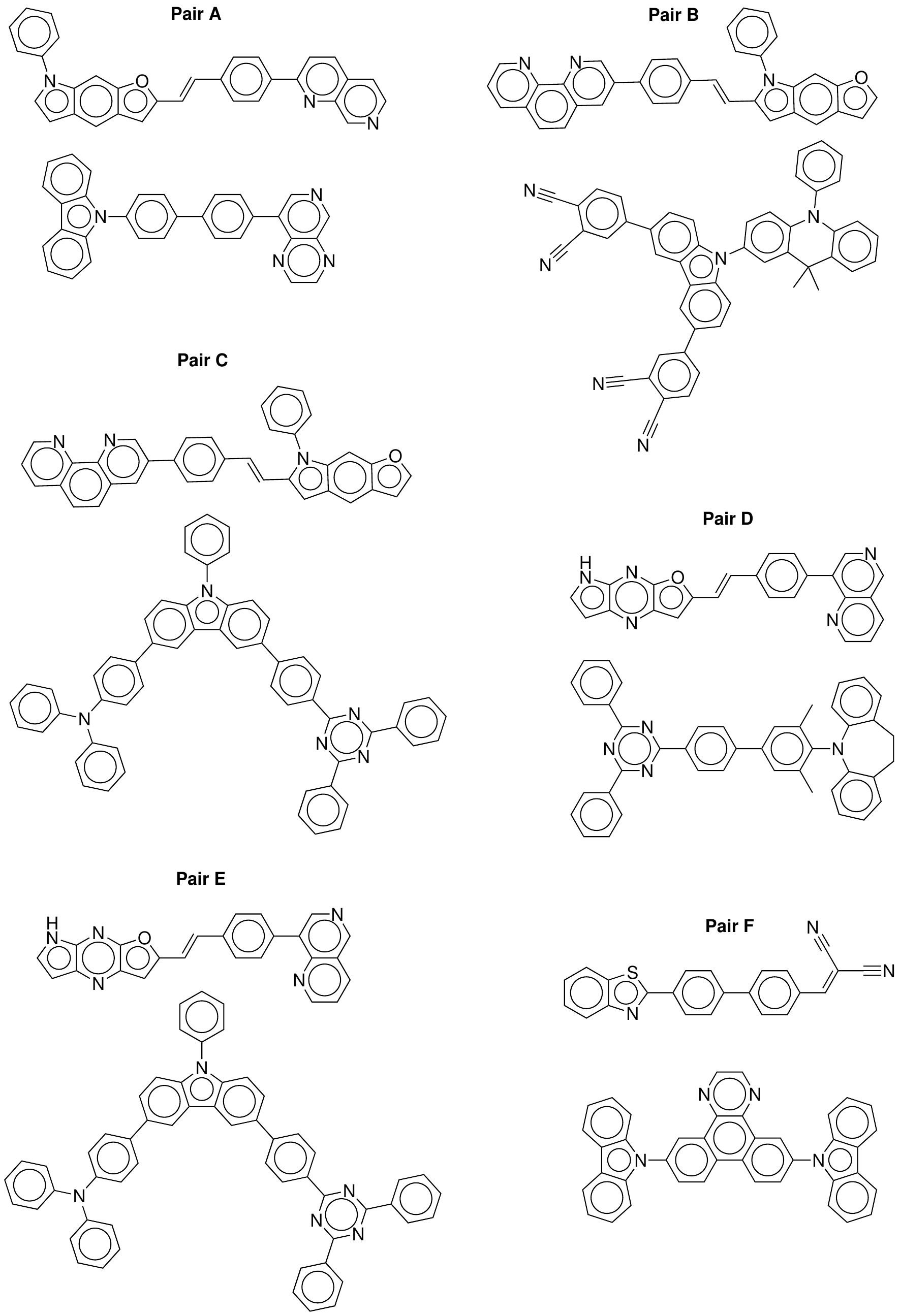}
\caption{\label{candidate pairs illustration}
Illustration of the bottom 6 pairs in Table.~\ref{pairs table}. The upper molecule of each pair corresponds to the optically active energy donor (molecule 2).
}
\end{figure} 

\section{The theoretical model}\label{Appendix B}

\subsection{Connection between model in main text and real systems}

The theoretical model in the main text considers a pair of two level systems in its representation of the molecular dimer. However, real molecules often possess a manifold of vibrational as well as higher excited states. As we explain in the following an effective two-level description nevertheless remains adequate for our current purpose, even if the constituent molecules possess a richer internal structure. The main reason for this is Kasha's rule, which applies due to the expected clear separation of timescales between vibrational relaxation (which happens on a fs timescale) and any of the much slower processes which are explicitly captured by our model.

Our model aims to find a general and effective yet justified description, that does not require a host of specific assumptions that will differ from dimer to dimer. As the following sections in this document show that our simplified model already displays rich dynamics and involves a number of subtleties. Naturally, its degree of validity will depend on the specifics of a molecular system at hand, but in all cases we expect it to provide valuable guiding insight as well as constituting a useful and adequate working description for a large number of molecular dimers. Once a particular candidate has been fully characterised, the model could easily be extended to account for the details in that particular scenario. 

We note that the initial photon absorption event in the cycle may in some systems produce vibrational as well as (delocalised) electronic excitation on the molecular dimer, but by Kasha's rule all excited state population quickly relaxes to the lowest excited equilibrium states of each molecule, which entirely determine the photoluminescence, {\it i.e.}~the optical emission properties. Provided this relaxation occurs fast enough, higher excited states can be adiabatically eliminated from the model, since they never carry any significant steady state population. We may therefore assume (incoherent) excitation directly into a mixture of population in the bright(er) $\ket{+} $ and the dark(er) $\ket{-}$ states. Ultimately, we desire to store all population in the dark state. Our model from the main text achieves this by means of a phonon-assisted process (due to the vibrational background) that is distinct from the vibrational relaxation of strongly coupled modes connected to Kasha's rule. Our model may therefore sometimes overestimate the amount of population that needs to be moved from the $\ket{+}$ to the $\ket{-}$ state, and when that is the case, it will be conservative in its estimate of the ensuing performance enhancement due to the presence of the dark state.

A second point relates to our estimate of the F\"orster coupling strength: F\"orster coupling is well understood in the regimes of weak and strong coupling. In the latter case -- arguably the more desirable one for our scheme -- the coupling strength is usually taken as a fixed coherent Hamiltonian contribution for modelling purposes. As we observe slight differences between absorption and emission properties of our candidate dimers (see Table~\ref{pairs table}) we cannot get a precise estimate of the coupling from the S0 transition. However, for our selected dimer candidates the large degree of similarity for the S0 and S1 transitions implies a sufficiently strong coupling near, or in between the two numbers shown. We note that the asymmetric approach in particular is remarkably robust to small variations in $V_F$ (also see Fig.~\ref{deviation from dark state}), so that small deviations away from predicated values would not greatly affect the achievable performance. A rigorous calculation of the coupling strength would be desirable but is a known tricky problem which would require a full calculation which explicitly considers the vibronic nature of  the excited states. This seems worthwhile of investigation on its own and is beyond the scope of the present paper. 

\subsection{Coupling to the Reaction Centre}\label{Coupling to the reaction centre}
In our model we use nonlocal electron-phonon coupling\cite{Munn1985,Munn1985a,Zhao1994,Mukamel2004} to capture the transfer into the reaction centre. We use this form because of its mathematical simplicity plus its ability to capture the suppression of the transfer $\ket{+} \rightarrow \ket{\alpha}$ as introduced in Ref.~\onlinecite{Creatore2013}. 

As this form is not as commonly used in the study of exciton transfer, we note that we could alternatively have implemented the reaction centre transfer with only local phonons. In that case, we would include the following additional coherent coupling term to the system Hamiltonian [Eq.~(1) in the main text] 
\begin{align}
\Hamiltonian_s \rightarrow & \Hamiltonian_s 
\\ & + \left(v_{1\alpha}\ket{1}\bra{\alpha} + v_{2\alpha}\ket{2}\bra{\alpha} + v_{\beta g}\ket{\beta}\bra{g} + H. c.\right),\notag
\end{align}
whilst substituting all of $\hat I_{1\alpha}, \hat I_{2\alpha}, \hat I_{\beta g}$ in Eq.~(3) for another local phonon bath $\hat I_{\beta \beta}$. This results in almost exactly the same final set of rate equations. The main difference is that there is no longer a complete cancellation of the $\ket{+} \rightarrow \ket{\alpha}$ transition in the symmetric model. The asymmetric model, which is the main focus of our paper, remains largely unaffected (there is an additional small rate which becomes negligible when the molecule is properly asymmetric). These slight differences in the rate equation model depend on the specifics of the reaction center (which we only model in an abstracted way anyway), and in particular they do not change any of our central conclusions. For this reason, we have adopted the nonlocal phonon model, which maps directly onto the approach used by Ref.~\onlinecite{Creatore2013}, for ease of comparison.

\subsection{Rate equation details}\label{Rate equation details}
The rate equations of the system are given by:
\begin{align}
\pd{}{t} P_+ =& -\gamma_{+-}[(N_{\omega_{+-}}+1)P_+-N_{\omega_{+-}}P_-]
\\ \notag &-\gamma_{+g}[(N^{T_h}_{\omega_{+g}}+1)P_+-N^{T_h}_{\omega_{+g}}P_g]
\\ \notag &-\gamma_{+\alpha}[(N_{\omega_{+\alpha}}+1)P_+-N_{\omega_{+\alpha}}P_g],
\\
\pd{}{t} P_- =& +\gamma_{+-}[(N_{\omega_{+-}}+1)P_+-N_{\omega_{+-}}P_-]
\\ \notag &-\gamma_{-g}[(N^{T_h}_{\omega_{-g}}+1)P_--N^{T_h}_{\omega_{-g}}P_g]
\\ \notag &-\gamma_{-\alpha}[(N_{\omega_{-\alpha}}+1)P_--N_{\omega_{-\alpha}}P_g],
\\
\pd{}{t} P_\alpha =& +\gamma_{+\alpha}[(N_{\omega_{+\alpha}}+1)P_+-N_{\omega_{+\alpha}}P_\alpha]
\\ \notag &+\gamma_{-\alpha}[(N_{\omega_{-\alpha}}+1)P_--N_{\omega_{-\alpha}}P_\alpha]
\\ \notag &-\gamma_{\alpha \beta}[(N_{\omega_{\alpha\beta}}+1)P_\alpha-N_{\omega_{\alpha\beta}}P_\alpha]
\\ \notag &-\chi \gamma_{\alpha \beta}[(N_{\omega_{\alpha g}}+1)P_\alpha-N_{\omega_{\alpha g}}P_g],
\\
\pd{}{t} P_\beta =& +\gamma_{\alpha \beta}[(N_{\omega_{\alpha\beta}}+1)P_\alpha-N_{\omega_{\alpha\beta}}P_\alpha]
\\ \notag & -\gamma_{\beta g}[(N_{\omega_{\beta g}}+1)P_\beta-N_{\omega_{\beta g}}P_g],
\end{align}
combined with the population normalisation condition $\sum_i P_i = 1$.
Here $\omega_{ab} = \epsilon_a-\epsilon_b$ and $N_\omega$ ($N^{T_h}_\omega$) is the thermal occupation number for a given frequency $\omega$ and temperature $T_c$ ($T_h$).
Assuming that the spectral densities of the environments are nearly flat around the transition frequencies, the different $\gamma$'s of the asymmetric model are given by
\begin{align}
\gamma_{+g} =& |z \avg{+|1} + \avg{+|2}|^2 \gamma_{2g},\\
\gamma_{-g} =& |z \avg{-|1} + \avg{-|2}|^2 \gamma_{2g},\\
\gamma_{+-} =& | \avg{+|1}|^2| \avg{-|1}|^2 (\gamma_{11}+ \gamma_{22}),\\
\gamma_{+\alpha} =& |\avg{+|1}|^2 \gamma_{1\alpha},\\
\gamma_{-\alpha} =& |\avg{-|1}|^2 \gamma_{1\alpha}.
\end{align}
When Eq.~(6) is satisfied (i.e.~we have a completely dark state), these reduce to 
\begin{align}
\gamma_{+g} =& \gamma_{1g}+\gamma_{2g}\\
\gamma_{-g} =& 0\\
\gamma_{+-} =& \frac{z^2}{(1+z^2)^2}(\gamma_{11}+\gamma_{22})\\
\gamma_{+\alpha} =& \frac{z^2}{1+z^2}\gamma_{1\alpha}\\
\gamma_{-\alpha} =& \frac{1}{1+z^2}\gamma_{1\alpha}.
\end{align}
We note that in the case where both molecules are coupled \emph{independently} to the reaction centre, {\it i.e.} $[\mu_{1\alpha}(t),\mu_{2\alpha}] = 0$, but with the same strength $\gamma_{1\alpha} = \gamma_{2\alpha}$, we get $\gamma_{-\alpha} = \gamma_{+\alpha} = \gamma_{1\alpha}$, regardless of the asymmetry. 

Please note that the above rate equation model is distinct from the most basic Pauli master equation treatment which simply discards coherences: our rate equations operate with respect to the diagonalised system basis, meaning coherence between the two chromophores is implicitly included in the model and plays an important part in the dynamics of the system.

\subsection{Current and Power plot}\label{Appendix IVPV}
\begin{figure}
\centering
\includegraphics[scale=0.75]{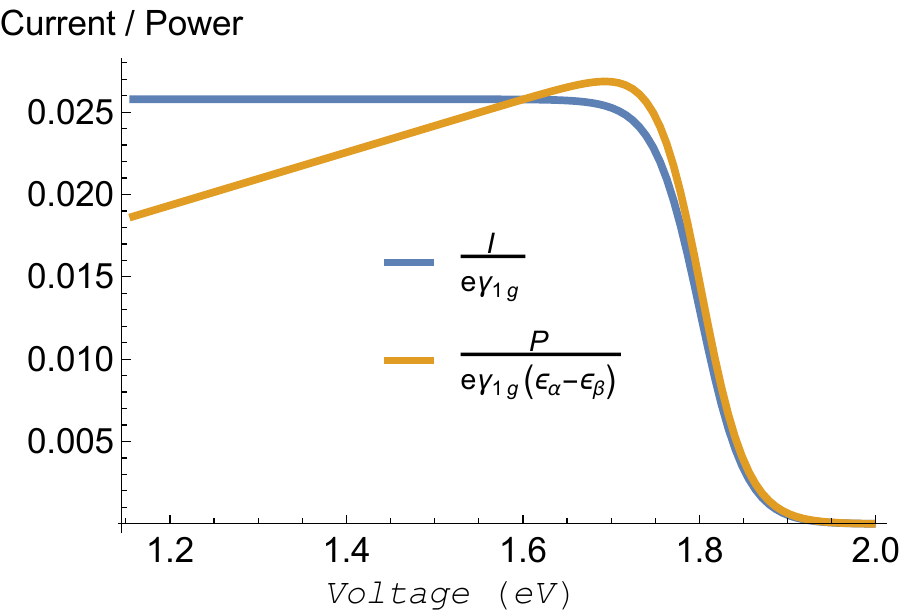}
\caption{\label{IVPVplot}
An example of a current/power plot as a function of voltage for the asymmetric model. 
The parameters are:
$\gamma_{1g}~+~\gamma_{2g}~=~1.24 \times 10^{-6}$~eV,
$\gamma_{1 \alpha }~=~6 \times 10^{-7}$~eV,
$\gamma_{11} = \gamma_{22} = 0.005$~eV,
$\gamma_{\beta g}~=~0.0248$~eV,
$T_h = 6000$~K, $T_c = 300$~K,
$\epsilon_- = 2$~eV,
$\epsilon_2-\epsilon_1 = 0.1$~eV
$J_{12} = 10$~meV
$\epsilon_\alpha = 1.8$~eV, $\epsilon_\beta = 0.2$~eV, $\chi = 0.2$.
}
\end{figure}

In Fig.~\ref{IVPVplot} we show a typical example of an I-V and P-V plot, given for the asymmetric model. This plot is produced by varying only the transfer rate inside the reaction centre $\gamma_{\alpha \beta}$, while fixing all other parameters. The power output of the cell is then given by the maximal power of this plot.

\subsection{Realistic imperfections }\label{Appendix deviation}
\begin{figure}
\centering
\includegraphics[scale=0.8]{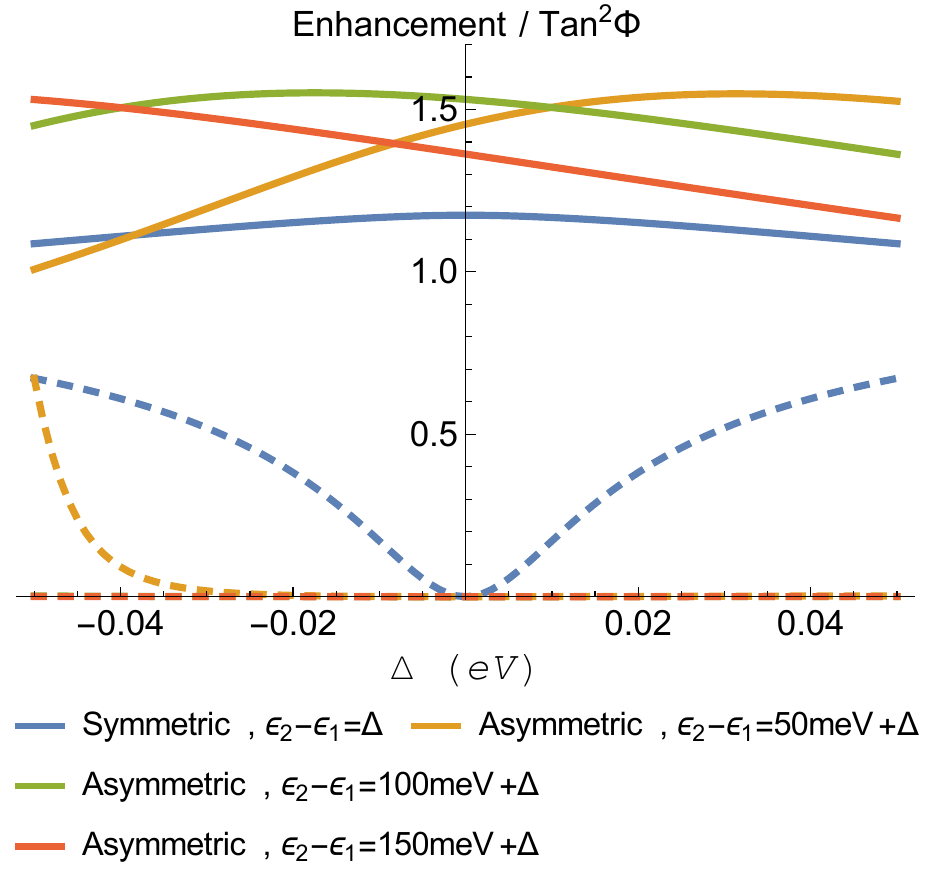}
\caption{\label{deviation from dark state}
A comparison of the relative power enhancement achievable with a deviation from the dark state, as a function of the deviation $\Delta$ of the energy difference $\epsilon_2-\epsilon_1$ from the ideal case where a dark state is present. The symmetric model and three different asymmetric model cases are shown. In solid is the relative enhancement of the symmetric and asymmetric models, and in dashed is $\tan^2\Phi$ for each model. 
Other parameters are inspired by Ref.~\onlinecite{Creatore2013}:
$\gamma_{1g} + \gamma_{2g} = 1.2 \times 10^{-6}$ eV,
$\gamma_{11} = \gamma_{22} = 0.005$ eV,
$\gamma_{1\alpha} = 6 \times 10^{-7}$ eV,
$\gamma_{\beta g} = 0.0248$ eV,
$T_h = 6000$~K, $T_c = 300$~K
$\epsilon_m = 2$ eV,
$J_{12} = 10$ meV,
$\epsilon_\alpha = 1.8$ eV, $\epsilon_\beta = 0.2$ eV, $\chi = 0.2$.
}
\end{figure}

In Fig.~\ref{deviation from dark state} we plot the relative enhancement of the symmetric and asymmetric models when deviating from the dark state. We examine several different $\epsilon_2-\epsilon_1$ values, for each we set $z$ value that satisfies Eq.~(6). Now adding a deviation from the dark state condition $\epsilon_2-\epsilon_1 \rightarrow \epsilon_2-\epsilon_1 + \Delta$, we plot the relative enhancement and the deviation from the dark state $\tan^2\Phi$. We find that the performance of the symmetric model is relatively robust over  deviations of tens of meV in the site energies, even-though its $\ket{-}$ state is then far from fully dark. We attribute this to the fact that for $J_{12} = 10$ meV there is a non-negligible rate $\ket{-}\rightarrow\ket{+}$, meaning population in the dark state is not fully protected even at $\Delta = 0$. Deviating from an ideal dark state thus only results  in minor corrections to the enhancement. We also find that for the asymmetric model, a deviation in the site energies only slightly shifts the system from the dark state ($\tan^2\Phi<0.05$). Interestingly, the performance generally increases either to the left or to the right of the $\Delta=0$ case, {\it i.e.}~when Eq~(6) is not strictly satisfied. Not only does this imply robustness against fluctuations in the site energies, it could also be exploited to design optimised asymmetric dimers away from the dark-state criterion. The reason for this surprising power enhancement at $\Delta \neq 0$ is due to several competing processes that happen when increasing (decreasing) $\Delta$: the transfer rate into the dark state $\gamma_{+-}$ decreases (increases), as well as the opposite transfer from the dark to the bright state, while the transfer from the dark state into the reaction centre $\gamma_{-\alpha}$, which is an important bottleneck, increases (decreases). Reducing any bottleneck at the expense of sacrificing some degree of dark-state protection leads to higher overall performance.

Another factor affecting the power output is the coupling to the reaction centre. For the symmetric model, following Ref.~\onlinecite{Creatore2013}, we assume that the two molecules are engineered in such a way that the two antenna chromophores couple to the reaction centre with a relative phase difference of $\pi$ (this could, for example, arise from the form of the relevant wave-function orbitals at site $\ket{\alpha}$). This particular choice renders the $\ket{+}$ state completely decoupled from the reaction centre, whereas the rate from the dark-state $\ket{-}$ to the reaction centre is maximal. Deviations from this idealised situation are to be expected when the molecules do not occupy exactly diametrically opposed sides of the reaction centre. In this case we can define an angle $\theta_{RC}$ that capture the phase difference between the coupling of the two molecules to the reaction centre via
\begin{gather}
\hat\mu_{2\alpha} = e^{i \theta_{RC}}\hat\mu_{1\alpha}.
\end{gather}

\begin{figure}
\centering
\includegraphics[scale=0.9]{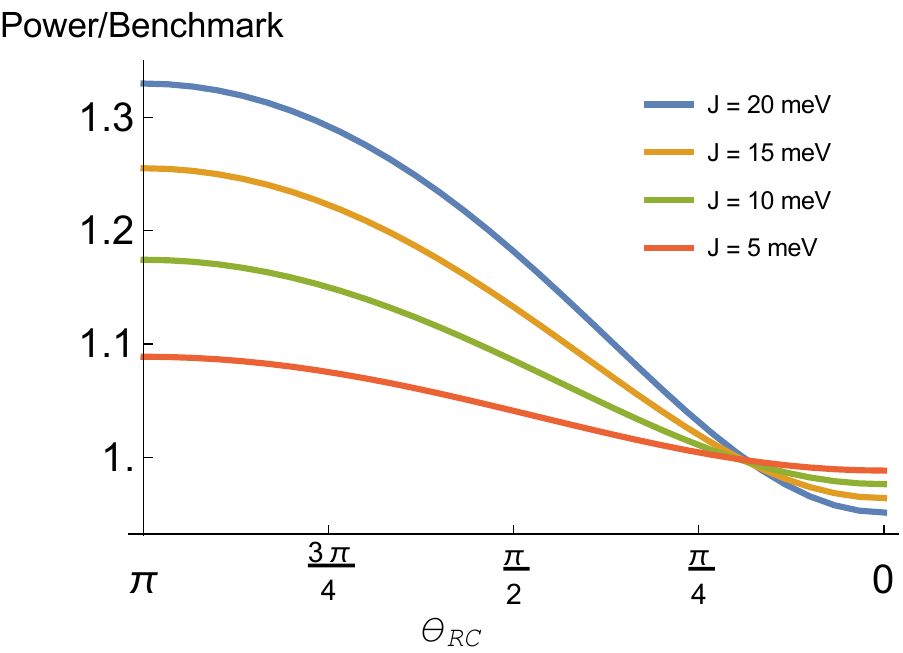}
\caption{\label{deviation from as state}
A plot of the power enhancement of the symmetric model when the coupling to the reaction centre is not perfectly antisymmetric.  
The parameters are inspired by Ref.~\onlinecite{Creatore2013}:
$\gamma_{1g} + \gamma_{2g} = 1.2 \times 10^{-6}$ eV,
$\gamma_{11} = \gamma_{22} = 0.005$ eV,
$\gamma_{1\alpha} = 6 \times 10^{-7}$ eV,
$\gamma_{\beta g} = 0.0248$ eV,
$T_h = 6000$~K, $T_c = 300$~K
$\epsilon_m = 1.8 eV$,
$\epsilon_\alpha = 1.8 eV$, $\epsilon_\beta = 0.2 eV$,$\chi = 0.2$.
}
\end{figure}

When $\theta_{RC} \neq \pi$, due to a different geometrical arrangement, static disorder or dynamical fluctuations, the transfer rates to the reaction centre are then given by:
\begin{align}
\gamma_{+\alpha} &= (1+J_{12}/\Omega_R \cos\theta_{RC})\gamma_{1\alpha},\\
\gamma_{-\alpha} &= (1-J_{12}/\Omega_R \cos\theta_{RC})\gamma_{1\alpha}.
\end{align}
Only for the case of no dipole mismatch ($\varphi=0$) and perfect antisymmetric coupling to the reaction centre ($\theta_{RC} = \pi$) do we obtain $\gamma_{+\alpha}/\gamma_{-\alpha} = \gamma_{-g}/\gamma_{+g} = \tan^2\Phi$ as assumed so far. Figure~\ref{deviation from as state} shows the relative enhancement of the symmetric model when the coupling to the reaction centre deviates from $\theta_{RC}=\pi$. We note that for small deviations (less than $\sim \pi/4$) the decrease in enhancement is only about 5\%.

Turning to the asymmetric system, only chromophore 1 is coupled to the reaction centre, so the concept of a relative phase difference does not arise. In this case, the transfer rates to the reaction centre are:
\begin{align}
\gamma_{+\alpha} &= \frac{1}{2}(1-(\epsilon_2-\epsilon_1)/\Omega_R)\gamma_{1\alpha}, \\
\gamma_{-\alpha} &= \frac{1}{2}(1+(\epsilon_2-\epsilon_1)/\Omega_R)\gamma_{1\alpha}. \label{asym rc rate}
\end{align}
An interesting observation here is that in the case of deviation from the dark state, the rate $\gamma_{-\alpha}$ can actually increase. As described above, this is one reason why the optimal set of parameters of the asymmetric dimer can be away from the fully dark state, as shown in Fig.~\ref{deviation from dark state}.

\subsection{Bloch-Redfield equation model}
The results shown in the main paper are based on a rate equation model. We here discuss the validity of this treatment, by comparing its results to those obtained from full Bloch-Redfield equations. As we demonstrate below, the rate equation model is more than adequate in the parameter regimes of interest. 

We begin by sketching a derivation of the Bloch-Redfield as well as the rate equations, both obtained from the second-order time convolutionless (TCL2) generator\cite{BreuerPetruccione2007}. The general TCL generator $\Kappa$ is a superoperator with the following interaction picture definition:
\begin{gather}
\pd{}{t} \rho(t) = \Kappa \rho(t) ~.
\end{gather}
For an interaction Hamiltonian of the form
\begin{align}
\Hamiltonian_I &= \eta\sum_\nu  V_\nu B_\nu\label{linear interaction} ~,
\end{align}
one can expand the TCL generator in powers of the interaction, $\Kappa = \sum_n \eta^n \Kappa_n$. For factorised initial conditions all odd powers vanish,~\cite{BreuerPetruccione2007} meaning the first non-vanishing contribution is from the second order term, given by
\begin{align}
\Kappa_2(t) = &-\sum_{\nu_0,\nu_1} \int_0^t dt_1 V_{\nu_0}^\times(t)R_{\nu_0,\nu_1}(t,t_1) ~. \label{Kappa2 full}
\end{align}
Here
\begin{align}
V_\nu^\times(t) \square \equiv & [V_\nu(t),\square]~,\\
V_\nu^\circ(t) \square \equiv & \{V_\nu(t),\square \}~,\\
R_{\nu_a,\nu_b}(t_a,t_b)  \equiv & D^{\nu_a,\nu_b}(t_a-t_b)V_{\nu_b}^\times(t_b)\notag\\&+i D^{\nu_a,\nu_b}_1(t_a-t_b)V_{\nu_b}^{\circ}(t_b)~,\label{R superOperator}
\end{align}
and $D^{\nu_a,\nu_b}(t),D^{\nu_a,\nu_b}_1(t)$ are the real and imaginary parts of the response function, respectively, given by 
\begin{align}\label{response matrix}
\alpha^{\nu_a,\nu_b}(t) &= D^{\nu_a,\nu_b}(t) + i D^{\nu_a,\nu_b}_1(t) \notag\\
& = \eta^2 \text{Tr}\Big\{B_{\nu_a}(t) B_{\nu_b}\rho_B \Big\}~.
\end{align}
Below we assume for simplicity that $\alpha^{\nu_a,\nu_b}(t) = \delta_{\nu_a,\nu_b}\alpha(t)$, i.e.~a single response function is given for independent baths. Generalisation to different response functions is straightforward.

It is useful to write $\Kappa_2$ in Liouville space, with $\ket{i}\bra{j} \rightarrow \ket{ij}$. In the system's eigenbasis the elements of $\Kappa_2$ are given by:
\begin{align}
\bra{i j} &\Kappa_2 \ket{rs} = -e^{-i(\Delta_{rs}-\Delta_{ij})t}\sum_\nu\int_0^t d\tau\Big\{
\notag\\ & \sum_k \Big[\delta_{js} e^{-i\Delta_{k r}\tau}V^\nu_{ik}V^\nu_{kr}\alpha(\tau) + \delta_{ir} e^{+i\Delta_{k s}\tau}V^\nu_{sk}V^\nu_{kj}\alpha^*(\tau) \Big]
\notag\\&-V^\nu_{ir}V^\nu_{sj}\Big[e^{-i\Delta_{ir}\tau}\alpha(\tau)+e^{+i\Delta_{js}\tau}\alpha^*(\tau)\Big]\Big\} ~,
\end{align}
where $\Delta_{ij} = \epsilon_i-\epsilon_j$, and $V^\nu_{ij} = \bra{i}V_\nu\ket{j}$.
At this point one can make the Markov approximation in order to simplify the equations. This includes extending the limit of the integral in the above equation to infinity. In our case, since we are looking for the steady-state of the system, this is not an approximation.
Now one can rewrite the TCL2 generator as
\begin{gather}
\bra{i j} \Kappa_2 \ket{rs} = e^{-i(\Delta_{rs}-\Delta_{ij})t} \bra{i j} \tilde\Kappa_2 \ket{rs} ~,
\end{gather}
where $\tilde{\Kappa}_2$ has no time-dependence. The master equation thus reads
\begin{gather}
\pd{}{t}\rho_{ij}(t) = \sum_{rs} e^{-i(\Delta_{rs}-\Delta_{ij})t} \bra{i j} \tilde\Kappa_2 \ket{rs} \rho_{rs}(t) ~,\label{Kappa2equation}
\end{gather}
reducing to the following expression for the populations
\begin{gather}
\pd{}{t}\rho_{ii}(t) = \sum_{rs} e^{-i\Delta_{rs}t} \bra{i i} \tilde\Kappa_2 \ket{rs} \rho_{rs}(t) ~.
\end{gather}

Following the Markov approximation one may perform the so-called `secular approximation', where oscillating terms in the generator proportional to $\exp\{-i (\Delta_{rs}-\Delta_{ij})t\}$ with $(\Delta_{rs}-\Delta_{ij}) \neq 0$, are assumed to average out due to a separation of timescales in the dynamics. This approximation, for the case where the system's energies are non-degenerate, \emph{decouples the coherences from the populations}, and one gets for the populations
\begin{gather}
\pd{}{t}\rho_{ii}(t) = \sum_{j} \bra{i i} \tilde\Kappa_2 \ket{jj} \rho_{jj}(t) ~.
\end{gather}
This is the rate equation model which use in the main text. We note that the above set of Equations, [i.e.~Eq. (4) in the main text] is then equivalent to, e.g., the Lindblad master equation from Ref.~\onlinecite{Rebentrost2009a} as far as the population dynamics is concerned. On the other hand, without performing the secular approximation, Eq.~(\ref{Kappa2equation}) reads upon transformation back to the Schr\"{o}dinger picture
\begin{gather}
\pd{}{t}\tilde{\rho}_{ij}(t) = -i\Delta_{ij}\tilde{\rho}_{ij}(t) + \sum_{rs} \bra{i j} \tilde\Kappa_2 \ket{rs} \tilde\rho_{rs}(t) ~.
\end{gather}
In the literature this is known as the Bloch-Redfield equation.

\begin{figure}
\centering
\includegraphics[scale=0.9]{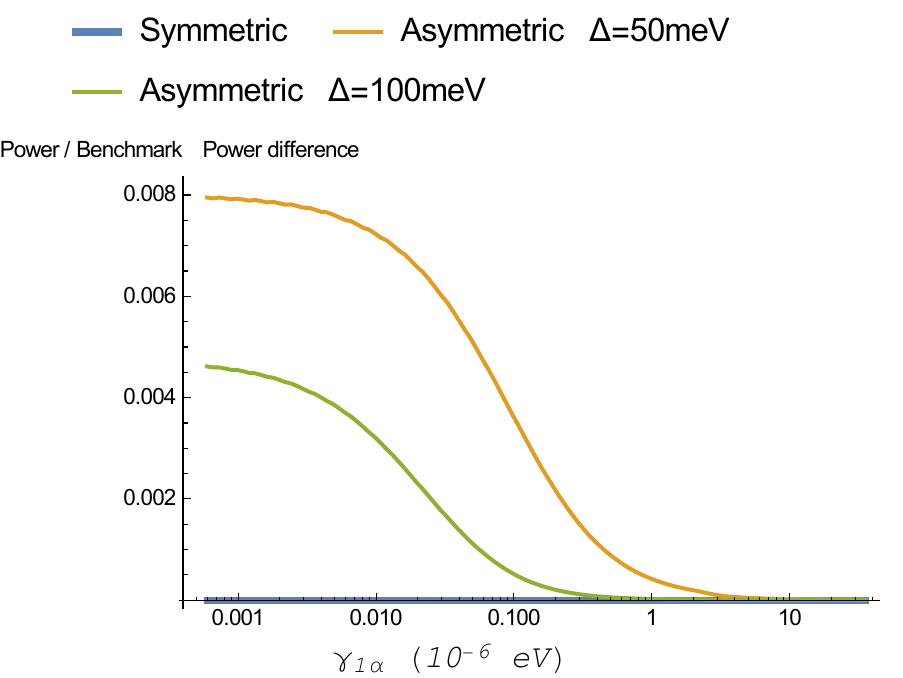}
\caption{\label{fig3_RedfieldDifference}
Difference between the enhancement given by the asymmetric and symmetric models with rate equations, and with the full Redfield theory. All parameters are the same as in Fig.~3 of the main text.
}
\end{figure}

\begin{figure}
\centering
\includegraphics[scale=0.6]{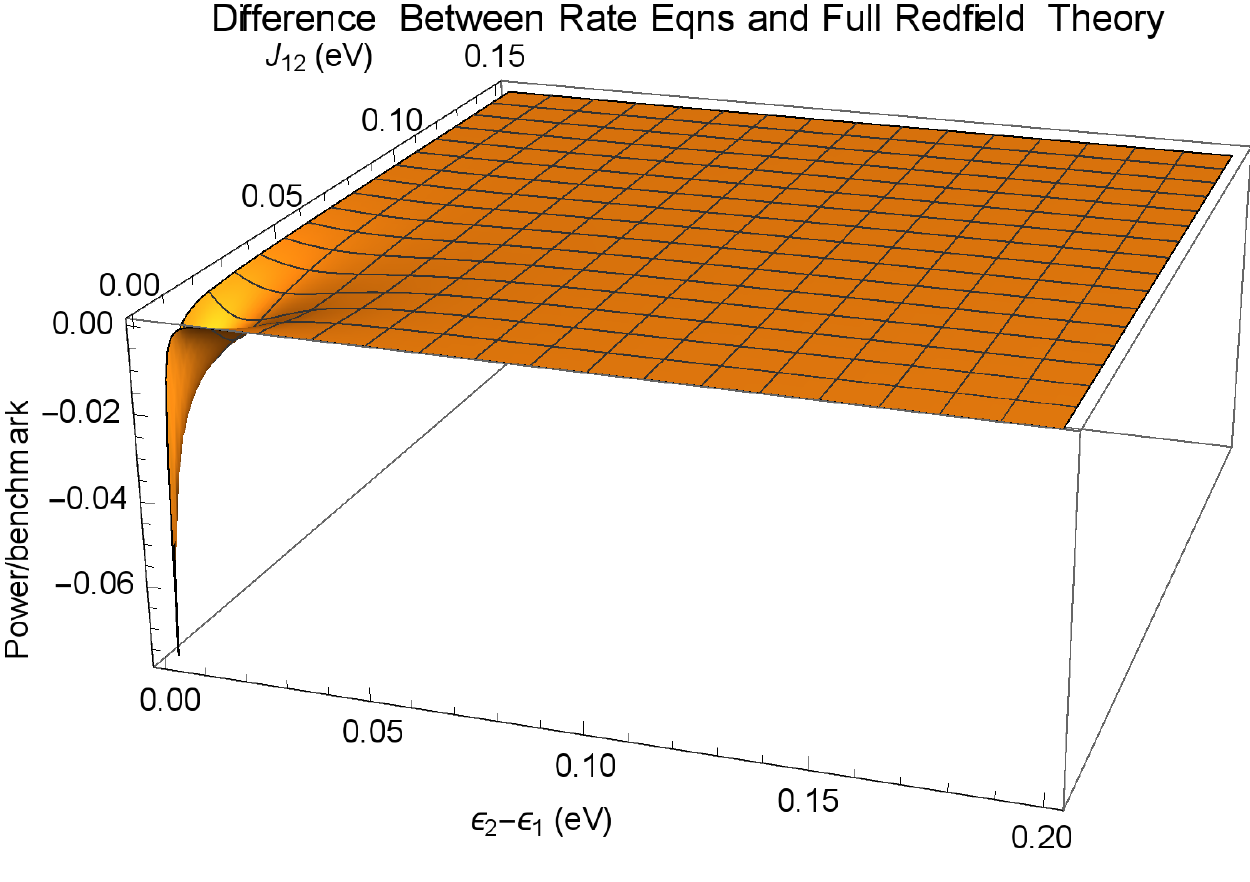}
\caption{\label{3dplot_RedfieldDifference}
Difference between the enhancement given by the asymmetric model with rate equations, and with the full Redfield theory. All parameters are the same as in Fig.~4 of the main text.
}
\end{figure}

To compare the full Bloch-Redfield treatment to our simplified rate equation model, we need to make additional assumptions about the imaginary parts of the TCL2 generator, which correspond to unitary renormalisation terms that do not feature in the rate equations. Expecting them to be small, we neglect (optical) Lamb shift terms whilst keeping (phonon) reorganisation energies, which we assume to be $10\%$ of the rates given by each bath, i.e. $\lambda_k = 0.1 \gamma_k$ (the precise choice is not important).
In Fig.~\ref{fig3_RedfieldDifference} we plot the difference between the power enhancement predicted by the rate equations and full Redfield theory, showing this for both the asymmetric and symmetric models, for the same parameters as in Fig.~3 of the main text. We find that in all cases the secular approximation is fully justified.
Figs.~\ref{3dplot_RedfieldDifference} shows this difference again this time for the same parameters as in Fig.~4 of the main text. We find that for the asymmetric case, the secular approximation is fully justified whenever the difference between the $\vert +\rangle$ and the $\vert - \rangle$ states is sufficiently large, which is the relevant regime for this work. For the symmetric case the two treatments coincide up to numerical inaccuracies of order $10^{-10}$ (not shown).

\subsection{Pure Dephasing}

\begin{figure}
\centering
\includegraphics[scale=0.8]{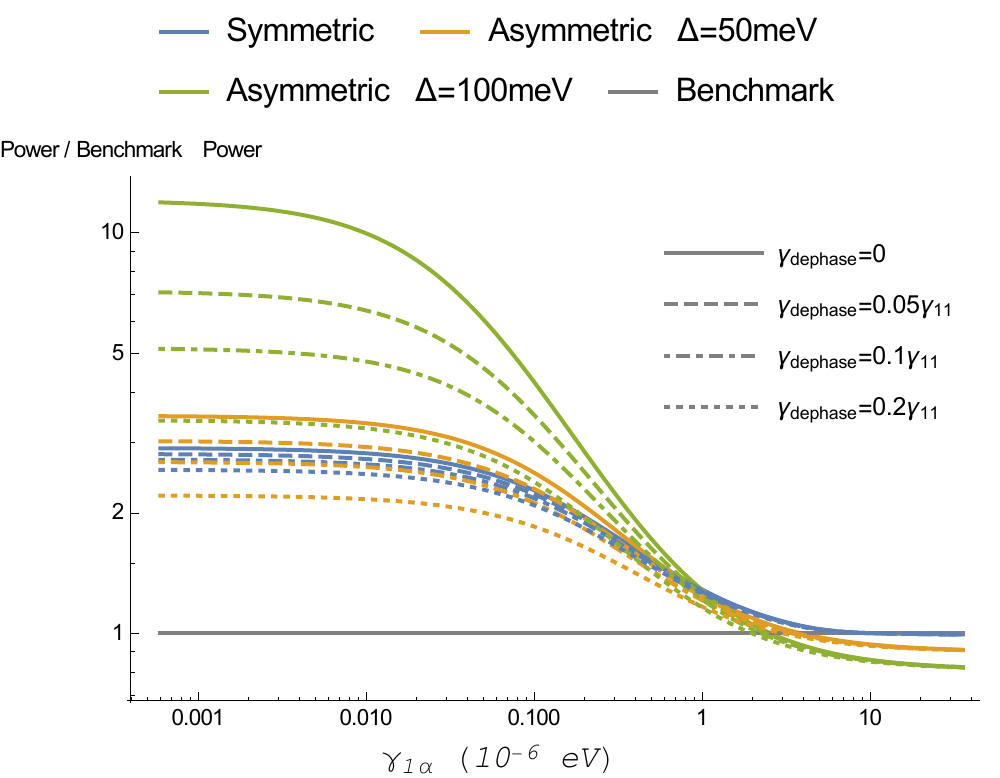}
\caption{\label{fig3_dephasing}
The enhancement given by the different models as a function of the rate into the reaction centre $\gamma_{1\alpha}$, with phenomological pure dephasing $\gamma_\text{dephase}$. All other parameters are the same as in Fig.~3 of the main text.
}
\end{figure} 

\begin{figure}
\centering
\includegraphics[scale=0.2]{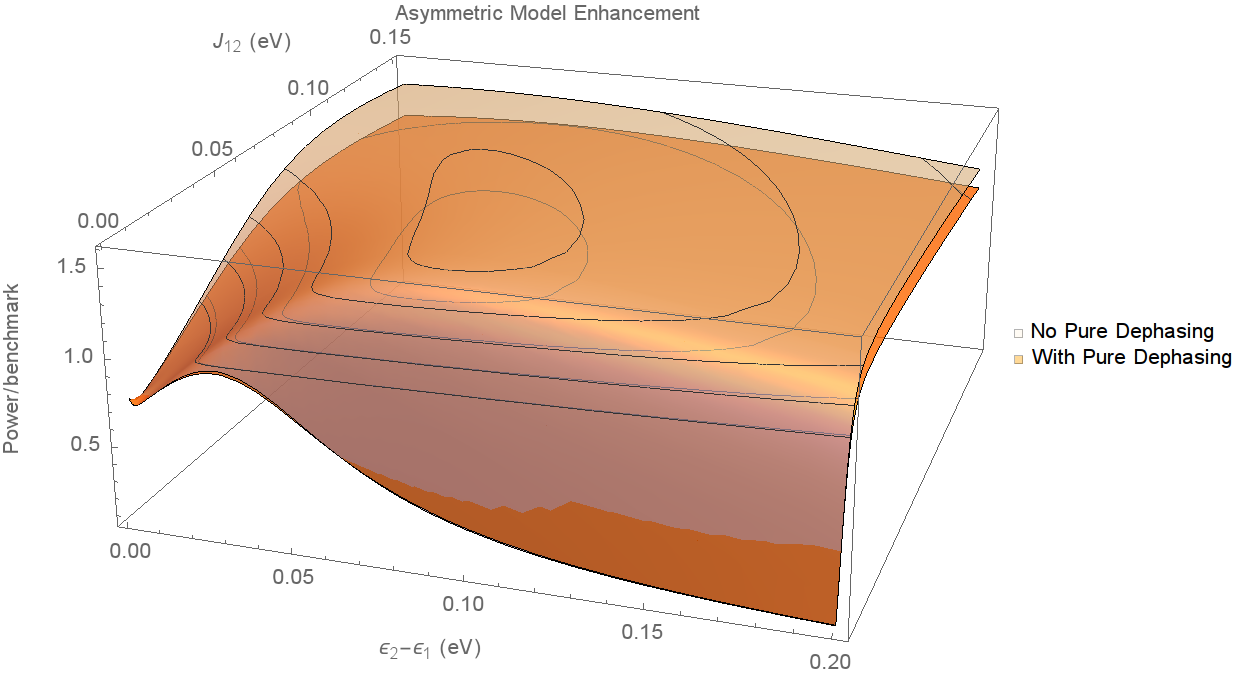}
\caption{\label{3dplot_dephasingAsymmetric}
The enhancement given by the asymmetric model with and without phenomological pure dephasing. Here, $\gamma_\text{dephase} = 0.1 \gamma_{11}$ and all other parameters are the same as in Fig.~4 of the main text.
}
\end{figure} 
\begin{figure}
\centering
\includegraphics[scale=0.8]{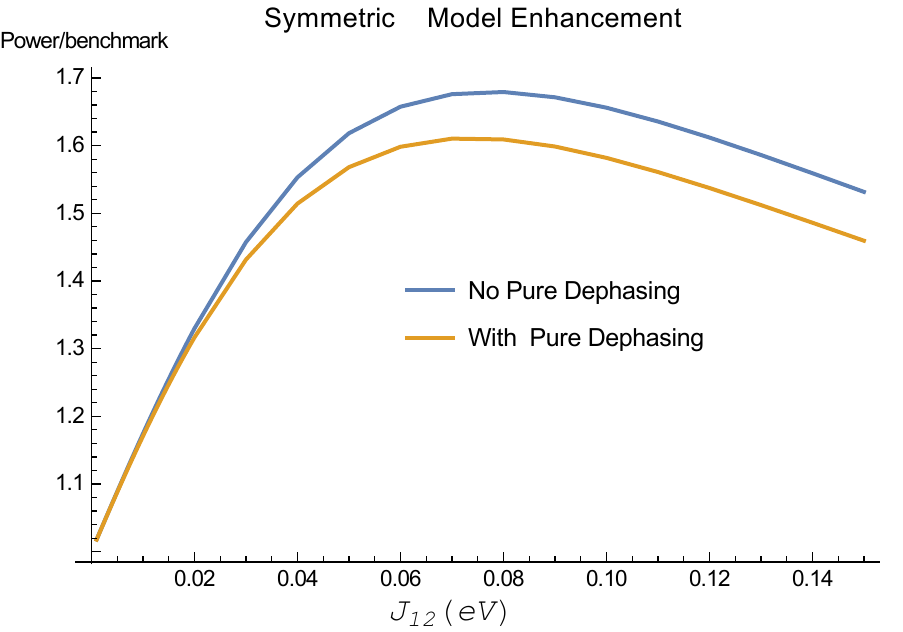}
\caption{\label{2dplot_dephasingSymmetric}
The enhancement given by the symmetric model with and without phenomological pure dephasing. Here, $\gamma_\text{dephase} = 0.1 \gamma_{11}$ and all other parameters are the same as in Fig.~4 of the main text.
}
\end{figure} 

We expect phonons to be the dominant source of decoherence and dephasing of our exciton states. Since their effect has already been accounted for based on a microscopic treatment, we here only need to consider small additional contributions, either due to other, unknown physical processes or inadequacies of our microscopic small, which does after all involve some approximations. Therefore, not being able to fall back on microscopically informed dephasing rates, we implement a phenomenological model (similar to the approach taken in, e.g., Ref.~\onlinecite{Rebentrost2009b}) to investigate how our results differ when pure dephasing terms defined in the site basis are included. Explicitly, we simply add a Lindblad operator to the Bloch-Redfield treatment, which is given by 
\begin{gather}
A_\text{d} = \sqrt{\gamma_\text{dephase}}\left(\ket{1}\bra{1} - \ket{2}\bra{2}\right)/\sqrt{2} ~,\\
\mathcal{L}_\text{dephase} \rho = A_\text{d} \rho A_\text{d} - 1/2 A_\text{d} A_\text{d} \rho - 1/2 \rho A_\text{d} A_\text{d} ~.
\\\notag
\end{gather}

As discussed above, we expect any additional pure dephasing rate to be much smaller than the level of decoherence that is already accounted for. Taking $\gamma_{11}$ as an indicative comparative measure, we will consider a dephasing rate that is 5-20\% of that.

Fig.~\ref{fig3_dephasing} shows the maximum relative enhancement of the different models, for the same parameters as in Fig.~3 of the main text. We find that the asymmetric model is very susceptible to pure dephasing, compared to the symmetric one, especially in the very weak coupling to the reaction centre regime. In some cases pure dephasing even brings the otherwise superior asymmetric model below the symmetric one. We believe this is because such pure dephasing, defined with respect to the site basis, effectively corresponds to a non-directional (or infinite temperature) transition $\ket{-} \leftrightarrow \ket{+}$, thus degrading the protection against exciton recombination given by the $\ket{-}$ state, which is the main advantage of the asymmetric model over the symmetric one.

Figs.~\ref{3dplot_dephasingAsymmetric} and \ref{2dplot_dephasingSymmetric} show the enhancements given by the asymmetric model and symmetric models, respectively, for the same parameters as in Fig.~4 of the main text. We find that for both cases, additional pure dephasing will reduce the enhancement by roughly 10\% for $\gamma_\text{dephase} = 0.1\gamma_{11}$. 
\vfill

\bibliography{bib}

\end{document}